\documentclass[sn-mathphys,Numbered]{sn-jnl}

\usepackage{graphics}
\usepackage{graphicx}%
\usepackage{amsmath,amssymb,amsfonts}%
\usepackage{amsthm}%
\usepackage{mathrsfs}%
\usepackage[title]{appendix}%
\usepackage{textcomp}%
\usepackage{manyfoot}%
\usepackage{booktabs}%
\usepackage{algorithm}%
\usepackage{algorithmicx}%
\usepackage{algpseudocode}%
\usepackage{listings}%

\usepackage{array}
\usepackage{multirow,bigdelim}
\usepackage{booktabs,tabu}
\usepackage{enumitem}

\usepackage{xcolor}
\definecolor{myblue}{rgb}{0,0.2,0.6}
\newcommand{\ignore}[1]{}
\usepackage{color}

\begin{document}

\title[Classification of large black holes]{Classification of \textquotedblleft large\textquotedblright  \ black holes into
seven families}
\author*[1]{Dafa Li}\email{lidafa@mail.tsinghua.edu.cn}
\author*[2]{Maggie Cheng} \email{maggie.cheng@iit.edu}
\author[3]{Xiongrong Li} \email{xli@math.uci.edu}
\author[2]{Shuwang Li} \email{sli15@iit.edu}




\affil*[1]{Department of Mathematical Sciences, Tsinghua University,
Beijing, 100084, China}

\affil[2]{Department of Applied Mathematics, Illinois Institute of
Technology, Chicago, IL 60616. USA}

\affil[3]{Department of Mathematics, University of California-Irvine,
Irvine, CA 92697, USA}


\abstract{
The black-hole--qubit correspondence has been proven to be ``useful for obtaining additional insight into
one of the string black hole theory and quantum information theory by exploiting approaches of the other"[Phys. Rev. D 82, 026003 (2010)].  Though different classes of stringy black holes can be related to the well-known stochastic local operations and classical communication (SLOCC) entanglement classes of pure states, the string theory requires a more detailed classification than the SLOCC classification of three qubits. In this paper, we derive the entanglement family of three qubits under local unitary operations (LU), and use the black-hole--qubit correspondence to classify \textquotedblleft large\textquotedblright\ black holes into seven inequivalent families. In particular, we show that two black holes with 4 non-vanishing charges ($q_{0}$, $p^{1}$, $p^{2}$, and $p^{3}$) are LU equivalent if their difference is only in the signs of charges. Thus, the classification of black holes is independent of the signs of charges and is only related to the ratio of the absolute values of charges. This observation simplifies the classification task, as one would only need to consider either the classification of non-BPS black holes or the classification of BPS black holes, but not both. Moreover, through the LU classification, the physical basis for this black-hole--qubit correspondence can be observed, and a relation between the black-hole entropy and the von Neumann entanglement entropy is revealed. Therefore, the LU classification offers a more straightforward physical connection than the SLOCC classification. Based on the LU classification, we further study the properties of von Neumann entanglement entropy for each of the seven families, and find the black holes with the maximal von Neumann entanglement entropy. 
}


\maketitle

\section{Introduction}
The entanglement classification of $n$ qubits under SLOCC was first proposed in \cite{Bennett}. Based on SLOCC classification, 
pure states of four qubits were classified into nine families \cite{Verstraete}, and pure states of three qubits were classified into six SLOCC equivalence classes: GHZ, W, AB-C, AC-B, BC-A, and A-B-C \cite{Dur}. The entanglement classification of three qubits under LU has also been studied. For example, in \cite{Acin00, Acin01}, pure states of three qubits were classified into five types: types 1, 2 (2a and 2b), 3 (3a and 3b), 4 (4a, 4b, and 4c), and 5. Further, the states of the GHZ SLOCC class were classified into types 2b, 3b, 4b, 4c, and 5 in \cite{Acin00,Acin01}.

In the seminal work \cite{Duff}, Duff first related quantum information theory to the physics of stringy black holes by expressing the entropy of the $STU$ black hole in terms of Cayley's hyperdeterminant of the
coefficients of a three-qubit pure state. Since then, the correspondence between the entanglement of qubits in quantum
information theory and black holes in string theory has been intensively studied \cite{Duff, Borsten09, Duff08, Linde, Levay, Levay07, Levay10, Borsten, Duff-talk, Broda, Geemen, Brown, Levay15}.

Kallosh and Linde \cite{Linde} showed that the entropy of the axion-dilaton extremal black hole is related to the concurrence of a two-qubit state. When the set of electric and magnetic charges is $(p^{0}$, $p^{1}$, $q_{0}$, $q_{1})$, the entropy of the axion-dilaton black hole is given by  $\frac{S}{\pi }=|p^{0}q_{1}-q_{0}p^{1}|$. It is also well known that the concurrence $C$ is $2|c_{0}c_{3}-c_{1}c_{2}|$ for a two-qubit state $|\psi \rangle=\sum_{i=0}^{3}c_{i}|i\rangle$ and the concurrence is the unique entanglement measure for pure states of two qubits. In \cite{Linde}, they indicated that if we identify the charges with the coefficients $c_{i}$ of  a
two-qubit state $|\psi \rangle $, i.e., $\left( 
\begin{array}{c}
p^{0} \\ 
p^{1} \\ 
q_{0} \\ 
q_{1}
\end{array}
\right) =\left( 
\begin{array}{c}
c_{0} \\ 
c_{1} \\ 
c_{2} \\ 
c_{3}
\end{array}
\right) $, then the entropy of axion-dilaton extremal black holes is
proportional to the concurrence of a two-qubit system, i.e., $S=\frac{\pi }{2}
C$. Thus, they established a correspondence between axion-dilaton extremal
black holes and two-qubit systems.

In  \cite{Linde, Kallosh-2021}, they investigated the relationship between the 3-tangle of a three-qubit state and the entropy of the $STU$ black holes, and related the well-known SLOCC entanglement classes of pure states of three qubits to different classes of black holes in string theory, and indicated that there are two types for black holes, i.e., large black holes and small black
holes. The large black holes with non-vanishing entropy correspond to states of GHZ SLOCC class, and small black holes with vanishing entropy correspond to states of the other five SLOCC classes. They also selected a representative for each SLOCC classification of extremal black holes. For example, they selected the black holes with the charges $q_{0}$, $p^{1}$, $p^{2}$, and $p^{3}$ as a representative for black holes corresponding to the GHZ SLOCC class. L\'{e}vay \cite{Levay} obtained a geometric classification of $STU$ black holes described in the language of twistor theory, and established a connection between the black hole entropy and the average real entanglement of the formation. Borsten et al. \cite{Borsten} derived the SLOCC classification of four-qubit entanglement invoking the black-hole--qubit correspondence.

The physics basis underpinning the connection between quantum entanglement and black holes is not well established as is pointed out in \cite{Duff, Borsten09, Levay}, although the Cayley's hyperdeterminant provides an interesting mathematical connection between stringy black hole entropy and quantum entanglement (specifically, 3-tangle). It is well known that the 3-tangle is different from the {von Neumann entropy of entanglement}. From the physics point of view, Cayley's hyperdeterminant in stringy black hole entropy and 3-tangle of three qubits remains a purely mathematical coincidence. It was also indicated in \cite{Linde} that the theory of stringy black holes requires a more detailed classification than the SLOCC classification of three qubits.

In this paper, we do a more detailed classification of GHZ SLOCC\ class under LU. In particular, we investigate the LU classification of black holes with charges $q_{0}$, $p^{1}$, $p^{2}$, and $p^{3}$, which correspond to GHZ SLOCC class. It is the first study to discuss LU classification of extremal black holes. Under the proposed LU classification, the black holes are classified into seven different families. We show that between two black holes with the aforementioned four non-vanishing charges, if the only difference is the signs of the charges, then they are LU equivalent. This means that the classification of black holes is not related to the signs of the charges. It is known that two LU equivalent states possess the same amount of entanglement and can be used to do the same tasks in quantum information theory \cite{Verstraete}. Therefore, LU classification provides a more straightforward physical connection. 

It is broadly recognized that von Neumann entropy has direct impact on our understanding of black holes and is an important means to describe black holes \cite{mat, kiran, geo, yi, bia}. For example, von Neumann entropy of black holes was studied \cite{geo} and a relationship between the Hawking radiation energy and von Neumann entropy in a conformal field emitted by a semiclassical two-dimensional black hole was found \cite{bia}, and the ratio of von Neumann entanglement entropy to the transverse growth of the exchanged surface is similar to the Bekenstein entropy ratio for a black-hole \cite{yi}. Therefore, in this paper we will also study the relation between the black hole entropy and the von Neumann entanglement entropy of three qubits. The relation between the two entropies contributes to the physical connection between quantum entanglement and black holes.

The paper is organized as follows. In Section \ref{sec:relation}, we review  the relation between stringy black holes and three-qubit states. In Section \ref{sec:lu-equivalence}, we establish the LU equivalency of two black holes differing only
by signs. In Section \ref{sec:7families}, we classify “large” black holes into seven families.
In Section \ref{sec:2entropy}, we discuss von Neumann entanglement entropy in quantum information theory and the entropy of black holes in string theory. In Section \ref{sec:maxvon}, we derive the black holes with the maximal von Neumann entanglement entropy. We give our conclusion in Section \ref{sec:summary}.

\subsection{Notations}
\begin{itemize}
\item $|\Psi \rangle$: denote a general pure state of three qubits. 
\item $|\psi \rangle$: denote the three-qubit state corresponding to a black hole with 4 non-vanishing charges $q_0$, $p^1$, $p^2$, and $p^3$, which belongs to the GHZ SLOCC class.
\item $\tau_{ABC}$: 3-tangle; $\tau_{AB}, \tau_{BC}, \tau_{AC}$: 2-tangles; $\tau_{A(BC)}$: denotes
the tangle between qubit $A$ and the pair $BC$, thinking of the pair $BC$ as a single object \cite{Coffman}.
\item $S(\rho_\mu)$: denote the von Neumann entanglement entropy.
\item $S_{bh}$: denote the classical supergravity entropy of black holes.
\item $q_j$: denote an integer electric charge.
\item $p^i$: denote an integer magnetic charge. The superscript $i$ is not an exponent; it is just an index.
\item $\lambda, \eta, \rho$: denote the coefficients of three qubits in the SD form.
\end{itemize}

\section{Relation between stringy black holes and three-qubit states}
\label{sec:relation}

The general static solution for a spherically symmetric black hole depends
on four integer electric charges, denoted as $q_{0}$, $q_{1}$, $q_{2}$, and $
q_{3}$, and four integer magnetic charges, denoted as $p^{0}$, $p^{1}$, $
p^{2}$, and $p^{3}$ \cite{Duff}. Note that the superscript $i$ in $p^{i}$ is not an
exponent; it is only an index. The STU black hole entropy $S_{bh}/\pi $  can be calculated via the 8
charges \cite{Behrndt, Linde}. We can simplify the expression for the
entropy as follows. Let

\begin{eqnarray}
\Delta  &=&(p^{0}q_{0}+p^{1}q_{1}+p^{2}q_{2}-p^{3}q_{3})^{2}  \notag \\
&&+4(p^{0}q_{3}-p^{1}p^{2})(p^{3}q_{0}+q_{2}q_{1}),  \label{entropy-1}
\end{eqnarray}%
then $(S_{bh}/\pi )^{2}=-\Delta $.

A three-qubit state in the Hilbert space involves eight terms. It can be written as $|\Psi \rangle =\sum_{\ell=0}^{7}a_{\ell }|\ell \rangle $ or $\sum_{i,j,k \in \{0,1\}}a_{ijk}|ijk\rangle $, where $a_l \in \mathbb{C}$.

Let $\det \Psi $ represent Cayley's hyperdeterminant of $|\Psi \rangle $, given as follows, 
\begin{eqnarray}
\det \Psi &=&(a_{0}a_{7}-a_{1}a_{6}-a_{2}a_{5}+a_{3}a_{4})^{2}  \notag \\
&&-4(a_{0}a_{3}-a_{1}a_{2})(a_{4}a_{7}-a_{5}a_{6}).  \label{Cayley}
\end{eqnarray}

It is known that the entanglement measure 3-tangle $\tau _{ABC}=4 \; |\det \Psi|$ (\cite{Miyake, Coffman}), where $|\cdot|$ denotes the absolute value.

To make a connection between the entropy of black holes and 3-tangle $\tau
_{ABC}$, it is necessary to make $\Delta =\det $ $\Psi $. When $\Delta =\det
\Psi $, for BPS and non-BPS, the classical supergravity entropy formula is
given as follows \cite{Linde}, 
\begin{equation}
S_{bh}=\frac{\pi }{2}\sqrt{\tau _{ABC}}.
\end{equation}%
To make $\Delta =\det $ $\Psi $, a dictionary between the eight charges for
a black hole and the eight coefficients of a three-qubit pure state is
needed. Three different dictionaries were given in the literature \cite%
{Duff, Linde, Levay}. Via eqs. \eqref{entropy-1} and \eqref{Cayley}, it is
easy to derive all the 16 dictionaries (see Appendix A).

In this paper, we will use the following dictionary \cite{Linde},

\begin{table}[!ht]
\caption{A dictionary between charges and coefficients.}
\label{tbl:dictionary}
\centering
\begin{tabular}{|l|c|c|c|c|c|c|c|c|}
\toprule Charges & $p^{0}$ & $p^{1}$ & $p^{2}$ & $p^{3}$ & $q_{0}$ & $q_{1}$
& $q_{2}$ & $q_{3}$ \\ 
\midrule Coefficients & $a_{0}$ & $-a_{1}$ & $-a_{2}$ & $-a_{4}$ & $a_{7}$ & 
$a_{6}$ & $a_{5}$ & $a_{3}$ \\ 
\bottomrule  
\end{tabular}
\end{table}

Kallosh and Linde \cite{Linde} investigated the black holes with four non-vanishing integer charges $q_{0}$, $p^{1}$, $p^{2}$, and $p^{3}$. In this paper, we only consider this kind of black holes. Thus, the three-qubit state corresponding to a black hole of this kind can be written as follows, 

\begin{eqnarray}
|\psi \rangle 
&=& p^0 |000\rangle  -p^1 |001\rangle -p^2 |010\rangle + q^3 |011\rangle -p^3 |100\rangle + q^2 |101\rangle + q_1 |110\rangle + q_0 |111\rangle \notag\\
&=&-p^{1}|001\rangle -p^{2}|010\rangle -p^{3}|100\rangle
+q_{0}|111\rangle .  \label{ghz-1}
\end{eqnarray}
The $|\psi \rangle $ in Eq. \eqref{ghz-1} belongs to GHZ SLOCC\ class (also called Mermin state) \cite%
{Linde, Duff08}. 

\vspace{5pt}
Let $\mu =1/\sqrt{(p^{1})^{2}+(p^{2})^{2}+(p^{3})^{2}+q_{0}^{2}}$, then $\mu |\psi \rangle $
becomes normalized. 

\vspace{5pt}

It is also easy to see that Cayley’s hyperdeterminant of $|\psi\rangle$ has the following property,
$$\det \psi =-4p^{1}p^{2}p^{3}q_{0}.$$

\vspace{5pt}

It was indicated in \cite{Duff-talk} that the 4-charge solution with just $%
q_{0}$, $p^{1}$, $p^{2}$, and $p^{3}$ may be considered as a bound state of
four individual black holes with charges $q_{0}$, $p^{1}$, $p^{2}$, and $%
p^{3}$, with zero binding energy.

The following state is a representative of real states of GHZ SLOCC class \cite{Duff08}.%
\begin{equation}
|\psi \rangle =-N_{3}|001\rangle -N_{2}|010\rangle -N_{1}|100\rangle
+N_{0}|111\rangle,
\label{eq:Ns}
\end{equation}
where $N_j \in \mathbb{N}$, for $j=0, 1 , 2, 3$.

It is known from \cite{Dur} that two pure states $|\psi \rangle
=\sum_{l=0}^{7}a_{l}|l\rangle $ and $|\psi ^{\prime }\rangle
=\sum_{l=0}^{7}a_{l}^{\prime }|l \rangle $ are LU equivalent if and
only if there are local unitary operators $\mathcal{A}$, $\mathcal{B}$, and $\mathcal{C}$
such that 
\begin{equation}
|\psi ^{\prime }\rangle = \mathcal{A} \otimes \mathcal{B} \otimes \mathcal{C} \; |\psi \rangle ,  \label{Dur}
\end{equation}

Accordingly, two STU black holes with charges $q_{0}$, $p^{1}, p^2,$ and $p^3$ are LU equivalent if and only if their corresponding states of three qubits are LU equivalent.

\section{LU equivalency of two black holes differing only by signs}
\label{sec:lu-equivalence}

The Schmidt decomposition (SD) for three qubits was proposed as follows \cite
{Acin00, Acin01}:  

\begin{equation}
\lambda _{0}|000\rangle +\lambda _{1}e^{i\varphi }|100\rangle +\lambda
_{2}|101\rangle +\lambda _{3}|110\rangle +\lambda _{4}|111\rangle ,
\label{Acin-1}
\end{equation}%
where $\lambda _{i}\geq 0$ for $i=0,1,2,3,4$, and $\sum_{i=0}^{4}\lambda _{i}^{2}=1$; $0\leq \varphi <2\pi $ is called the phase. Note that SD is normalized by definition.

\subsection{Two non-BPS black holes with $p^{1}p^{2}p^{3}q_{0}<0$ differing
only by signs are LU equivalent}

We show that two non-BPS black holes with $p^{1}p^{2}p^{3}q_{0}<0$ differing
only by signs are LU equivalent. For example, for a black hole with $p^{1}=p^{2}=p^{3}=1$, $q_{0}=-1$, and the
other one with $p^{1}=1$, $p^{2}=p^{3}=q_{0}=-1$, their corresponding states are LU equivalent.

For two black holes with $p^{1}p^{2}p^{3}q_{0}<0$, there are eight cases: 
\begin{itemize}
\item $S_{1}=\{p^{1}<0, p^{2}>0, p^{3}>0, q_{0}>0\}$,

\item $S_{2}=\{p^{1}>0, p^{2}<0, p^{3}>0, q_{0}>0\}$,

\item $S_{3}=\{p^{1}>0, p^{2}>0, p^{3}<0, q_{0}>0\}$,

\item $S_{4}=\{p^{1}>0, p^{2}>0, p^{3}>0, q_{0}<0\}$,

\item The mirror cases $S_{i}^{\prime }$ of $S_i$ for $i=1,2,3,4$, in which all
charges flip their signs.
\end{itemize}

\vspace{5pt}

We first calculate the Schmidt decomposition of $\mu |\psi \rangle $ in Eq. \eqref{ghz-1} with $%
p^{1}p^{2}p^{3}q_{0}<0$, then we show that for $S_{i}$ and $S_{i}^{\prime }$%
, their SDs are the same. 

\subsubsection{Schmidt decomposition of $\protect\mu |\protect\psi \rangle $ in Eq. 
\eqref{ghz-1} with $p^{1}p^{2}p^{3}q_{0}<0$}

To find the SD of $\mu |\psi \rangle $ in Eq. \eqref{ghz-1}, we need to
construct three unitary operators $U^{A}$, $U^{B}$, and $U^{C}$, which are
applied to the qubits $A$, $B$, and $C$, respectively.

Let $t=\frac{u_{00}^{A}}{u_{01}^{A}}$. By solving Eq. (8) in \cite{dli-jpa}, we have, 
\begin{equation}
t^{2}=-\frac{p^{3}q_{0}}{p^{1}p^{2}}.
\end{equation}
For $p^{1}p^{2}p^{3}q_{0}<0$, we have $t^2>0$. Then, a tedious calculation yields

\begin{eqnarray}
U^{A} &=&\frac{1}{\sqrt{|t|^{2}+1}}\left( 
\begin{array}{cc}
t & 1 \\ 
1 & -t%
\end{array}%
\right) , \\
U^{B} &=&\frac{1}{\sqrt{|k|^{2}+1}}\left( 
\begin{array}{cc}
1 & k \\ 
k & -1%
\end{array}%
\right) , \\
U^{C} &=&\frac{1}{\sqrt{a^{2}+b^{2}}\sqrt{t^{2}+1}}\left( 
\begin{array}{cc}
-p^{3} & -tp^{1} \\ 
-tp^{1} & p^{3}%
\end{array}%
\right) ,
\end{eqnarray}

where 
\begin{eqnarray}
t &=&\sqrt{-\frac{p^{3}q_{0}}{p^{1}p^{2}}}=\sqrt{\frac{|p^{3}q_{0}|}{%
|p^{1}p^{2}|}}, \notag \\
k &=&\frac{tp^{2}}{p^{3}}=-\frac{q_{0}}{tp^{1}},  \notag \\
a &=&-p^{3}/\sqrt{t^{2}+1}, \notag \\
b &=&-tp^{1}/\sqrt{t^{2}+1}.  \notag
\end{eqnarray}

Note that $t,k,a,$ and $b$ are all real. Let 
\begin{equation}
\chi =\frac{1}{\sqrt{k^{2}+1}\sqrt{a^{2}+b^{2}}(t^{2}+1)}.
\end{equation}%

A straightforward calculation of the $|\psi'\rangle$ in \eqref{Dur} yields:
\begin{eqnarray}
|\psi'\rangle &=&U^{A}\otimes U^{B}\otimes U^{C}|\psi \rangle \\
&=&\eta _{0}|000\rangle +\eta _{1}|100\rangle +\eta _{2}|101\rangle +\eta
_{3}|110\rangle +\eta _{4}|111\rangle,  \notag
\end{eqnarray}%
where  
\begin{eqnarray}
\eta _{0} &=&\sqrt{a^{2}+b^{2}}\sqrt{k^{2}+1}, \\
\eta _{1} &=&\chi t \left[(p^{1})^{2}+(p^{2})^{2}-(p^{3})^{2}-q_{0}^{2}%
\right],  \label{ada-1} \\
\eta _{2} &=&-\chi \left( t^{2}+1\right) \left( p^{1}p^{3}+p^{2}q_{0}\right)
,  \label{ada-2} \\
\eta _{3} &=&-\chi \left( t^{2}+1\right) \left( p^{2}p^{3}+p^{1}q_{0}\right)
,  \label{ada-3} \\
\eta _{4} &=&-2 \chi t \left( p^{1}p^{2}-p^{3}q_{0}\right) .
\end{eqnarray}

Clearly, all coefficients $\eta _{j}$ are real, and $\eta _{0}>0$. We can write $\eta
_{j}=|\eta _{j}|e^{i\theta _{j}}$ for $j=1,2,3,4$, where $\theta _{j}=0$ or $
\pi $. Then, $|\psi \rangle $ becomes the following using the method in \cite
{dli-jpa}: 

\begin{eqnarray}
|\psi ^{\prime }\rangle =\eta _{0}|000\rangle +|\eta _{1}|e^{i\phi
}|100\rangle +|\eta _{2}||101\rangle  + |\eta _{3}||110\rangle +|\eta _{4}||111\rangle ,  \label{ASD-1-}
\end{eqnarray}
where $\phi =\theta _{1}-\theta _{2}-\theta _{3}+\theta _{4}$. 
Since $\theta _{j}=0$ or $\pi $, it is easy to see $e^{i\phi }=\pm 1$. Therefore, $|\psi ^{\prime }\rangle $ is a real
state.

Computing the normalizing factor $\mu$ from \eqref{ASD-1-}, we obtain $\mu |\psi ^{\prime }\rangle $, which is the SD of $\mu |\psi \rangle $ in Eq. \eqref{ghz-1}.

\subsubsection{Two non-BPS black holes with $p^{1}p^{2}p^{3}q_{0}<0$ differing only by signs are LU equivalent}

In the following we establish that two non-BPS black holes with $p^{1}p^{2}p^{3}q_{0}<0$ differing only by signs are LU equivalent. Consider the $|\psi'\rangle$ in \eqref{ASD-1-}. We want to show that all eight cases ($S_{i}$ and $S_{i}^{\prime }$, $
i=1,2,3,4$) have the same $|\psi'\rangle$.

First, we calculate $|\eta _{j}|$. For $\eta _{2}$ in Eq. \eqref{ada-2}, $
p^{1}p^{3}$ and $p^{2}q_{0}$ have the opposite signs because $%
p^{1}p^{2}p^{3}q_{0}<0$. Thus, $|\eta _{2}|=\chi \left( t^{2}+1\right)
||p^{1}p^{3}|-|p^{2}q_{0}||$. Similarly, $|\eta _{3}|=\chi \left(
t^{2}+1\right) ||p^{2}p^{3}|-|p^{1}q_{0}||$ and $|\eta _{4}|=2\chi
t(|p^{1}p^{2}|+|p^{3}q_{0}|)\neq 0$. Clearly, $|\eta _{j}|$, $j=0,1,2,3,4$, are independent of the sign of
charges. Therefore, $|\eta _{j}|$ is the same for $S_{i}$ and 
$S_{i}^{\prime }$, $i=1,2,3,4$.

Next, we show $e^{i\phi}$ are the same for $S_{i}$ and $S_{i}^{\prime }$, $i=1,2,3,4$. Since $S_{i}^{\prime }$ are mirror to $S_{i}$, we only need to consider $S_{i}$ below.

From Eq. \eqref{ada-1}, clearly $\theta _{1}$ is the same for $S_{i}$, $i=1,2,3,4$.

There are four cases from eqs. \eqref{ada-2} and \eqref{ada-3} based on the
values of $\eta_2$ and $\eta_3$:

\begin{itemize}
\item A). $|p^{2}p^{3}|>|p^{1}q_{0}|$ and $|p^{1}p^{3}|>|p^{2}q_{0}|$;

\item B). $|p^{2}p^{3}|>|p^{1}q_{0}|$ and $|p^{1}p^{3}|<|p^{2}q_{0}|$;

\item C). $|p^{2}p^{3}|<|p^{1}q_{0}|$ and $|p^{1}p^{3}|>|p^{2}q_{0}|$;

\item D). $|p^{2}p^{3}|<|p^{1}q_{0}|$ and $|p^{1}p^{3}|<|p^{2}q_{0}|$.
\end{itemize}

For each case, we can see that $e^{i\phi }$ is the same for $S_{i}$, $
i=1,2,3,4$.

Thus, for all eight cases $S_{i}$ and $S_{i}^{\prime }$, $i=1,2,3,4$,  $|\psi
\rangle $ in Eq. \eqref{ghz-1} is transformed into the same $|\psi ^{\prime
}\rangle $ under LU. Thus, all $|\psi \rangle $ with $p^{1}p^{2}p^{3}q_{0}<0$
in Eq. \eqref{ghz-1} are LU equivalent \cite{dli-jpa}. For example, one black hole with $\ p^{1}=\ p^{2}=\ p^{3}=1$ and $\ q_{0}=-1$
, the other one with $\ p^{1}=\ p^{2}=1$, $\ p^{3}=-1$ and $\ q_{0}=1$, their $
|\psi ^{\prime }\rangle $ are both $\sqrt{2} (|000\rangle +|111\rangle )$.

\subsection{ Two BPS black holes with $p^{1}p^{2}p^{3}q_{0}>0$ differing
only by signs are LU equivalent}

For example, one with $p^{1}=p^{2}=1$, $p^{3}=q_{0}=-1$, and the other one
with $p^{2}=p^{3}=-1$, $p^{1}=q_{0}=1$, we can show that their corresponding
states are LU equivalent.

Next, generally consider two black holes with $p^{1}p^{2}p^{3}q_{0}>0$.

For $p^{1}p^{2}p^{3}q_{0}>0$, there are eight cases:

\begin{itemize}
\item $H_{1}=\{p^{1}>0,p^{2}>0,p^{3}<0,q_{0}<0\},$

\item $H_{2}=\{p^{1}>0,p^{2}<0,p^{3}>0,q_{0}<0\},$

\item $H_{3}=\{p^{1}>0,p^{2}<0,p^{3}<0,q_{0}>0\},$

\item $H_{4}=\{p^{1}>0,p^{2}>0,p^{3}>0,q_{0}>0\},$

\item The mirror cases $H_{i}^\prime$ of $H_i$, for $i=1,2,3,4$, in
which all charges flip their signs.
\end{itemize}

We first calculate the SD of $\mu |\psi \rangle $ in Eq. \eqref{ghz-1} with $%
p^{1}p^{2}p^{3}q_{0}>0$, then we show that for $H_{i}$ and $H_{i}^{\prime }$%
, their SDs are the same.

\subsubsection{Schmidt decomposition of $\protect\mu |\protect\psi \rangle $ in Eq. 
\eqref{ghz-1} with $p^{1}p^{2}p^{3}q_{0}>0$}

To find SD of $\mu |\psi \rangle $ in Eq. \eqref{ghz-1}, we need to
construct three unitary operators $W^{A}$, $W^{B}$, and $W^{C}$, which are
applied to the qubits $A$, $B$, and $C$, respectively. 

Let $\tilde{t}=\frac{w_{00}^{A}}{w_{01}^{A}}$%
. By solving Eq. (8) in \cite{dli-jpa}, we obtain 
\begin{equation}
\tilde{t}^{2}=-\frac{p^{3}q_{0}}{p^{1}p^{2}}.
\end{equation}

For $p^{1}p^{2}p^{3}q_{0}>0$, we have $\tilde{t}^{2}<0$. A tedious calculation
produces 
\begin{eqnarray}
W^{A} &=&\frac{1}{\sqrt{|\tilde{t}|^{2}+1}}\left( 
\begin{array}{cc}
\tilde{t} & 1 \\ 
1 & -\tilde{t}^{\ast }%
\end{array}%
\right) , \\
W^{B} &=&\frac{1}{\sqrt{|\tilde{k}|^{2}+1}}\left( 
\begin{array}{cc}
1 & \tilde{k}^{\ast } \\ 
\tilde{k} & -1%
\end{array}%
\right) , \\
W^{C} &=&\frac{1}{\sqrt{\tilde{a}^{2}+|\tilde{b}|^{2}}\sqrt{|\tilde{t}|^{2}+1%
}}\left( 
\begin{array}{cc}
-p^{3} & -\tilde{t}^{\ast }p^{1} \\ 
-\tilde{t}^{\ast }p^{1} & -p^{3}%
\end{array}%
\right),  \notag \\
\end{eqnarray}%
where $\tilde{t}^{\ast }$ is the conjugate of $\tilde{t}$, and

\begin{eqnarray}
\tilde{t} &=&i\sqrt{\frac{p^{3}q_{0}}{p^{1}p^{2}}}=i\sqrt{\frac{|p^{3}q_{0}|
}{|p^{1}p^{2}|}}, \notag \\
\tilde{k} &=&\frac{\tilde{t}p^{2}}{p^{3}}=-\frac{q_{0}}{\tilde{
t}p^{1}},  \notag \\
\tilde{a} &=&-p^{3}/\sqrt{|\tilde{t}|^{2}+1}, \notag \\
\tilde{b} &=& -\tilde{t}p^{1}/\sqrt{
|\tilde{t}|^{2}+1},  \notag
\end{eqnarray}%

where $\tilde{a}$ is a real number; $\tilde{t}$, $\tilde{k}$ and $\tilde{b}$ are imaginary numbers. Let 
\begin{equation}
\tilde{\chi}=\frac{1}{\sqrt{|\tilde{k}|^{2}+1}\sqrt{\tilde{a}^{2}+|\tilde{b}%
|^{2}}(|\tilde{t}|^{2}+1)}.
\end{equation}%
Then, a straightforward calculation yields

\begin{eqnarray}
|\psi'' \rangle &=&  W^{A}\otimes W^{B}\otimes W^{C}|\psi \rangle  \notag \\
&=&\rho _{0}|000\rangle +\rho_{1}|100\rangle +\rho _{2}|101\rangle +\rho _{3}|110\rangle +\rho _{4}|111\rangle ,
\label{lu-2}
\end{eqnarray}
where 
\begin{eqnarray}
\rho _{0} &=&\sqrt{|\tilde{a}|^{2}+|\tilde{b}|^{2}}\sqrt{|\tilde{k}|^{2}+1},
\\
\rho _{1} &=&\tilde{\chi}\tilde{t}^{\ast }\left(
(p^{1})^{2}+(p^{2})^{2}-(p^{3})^{2}-q_{0}^{2}\right) ,  \label{ru-1} \\
\rho _{2} &=&\tilde{\chi}\left( |\tilde{t}|^{2}+1\right) \left(
p^{1}p^{3}-p^{2}q_{0}\right) ,  \label{ru-2} \\
\rho _{3} &=&-\tilde{\chi}\left( |\tilde{t}|^{2}+1\right)
(p^{2}p^{3}-p^{1}q_{0}),  \label{ru-3} \\
\rho _{4} &=&2\tilde{\chi}\tilde{t}(p^{1}p^{2}+p^{3}q_{0}).
\end{eqnarray}

Clearly, $\rho _{0}>0$ is real, $\rho _{2}$ and $\rho _{3}$ are real; $\rho _{1}$
and $\rho _{4}$ are imaginary numbers. We can write $\rho _{j}=\left\vert
\rho _{j}\right\vert e^{i\omega _{j}}$, for $j=1,2,3,4$, where $\omega _{1}=\pm
\pi /2$, $\omega _{2}= 0 \text{ or } \pi $, $\omega _{3}= 0 \text{ or } \pi $, and $\omega _{4}=\pm \pi /2$.

Then, $|\psi \rangle $\ becomes the following by using the method in \cite{dli-jpa}:
\begin{eqnarray}
|\psi ^{\prime \prime }\rangle =\rho _{0}|000\rangle +|\rho_{1}|e^{i\varphi }|100\rangle +|\rho _{2}||101\rangle +|\rho _{3}||110\rangle +|\rho _{4}||111\rangle ,  \label{ASD-2-1}
\end{eqnarray}%
where $\varphi =\omega _{1}-\omega _{2}-\omega _{3}+\omega _{4}$. It is easy
to see $e^{i\varphi }=\pm 1$, thus $|\psi ^{\prime \prime }\rangle $ is a real
state. Normalizing it, we obtain the SD $\mu |\psi ^{\prime \prime }\rangle $.

Next, we calculate $|\rho _{2}|$. For $\rho _{2}$ in Eq. \eqref{ru-2}, $%
p^{1}p^{3}$ and $p^{2}q_{0}$ have the same sign because $%
p^{1}p^{2}p^{3}q_{0}>0$. Therefore, $|\rho _{2}|=\tilde{\chi}\left( |\tilde{t%
}|^{2}+1\right) ||p^{1}p^{3}|-|p^{2}q_{0}||$. Similarly, $|\rho _{3}|=\tilde{%
\chi}\left( |\tilde{t}|^{2}+1\right) ||p^{2}p^{3}|-|p^{1}q_{0}||$ and $|\rho
_{4}|=2\tilde{\chi}|\tilde{t}|(|p^{1}p^{2}|+|p^{3}q_{0}|)\neq 0$.

For example, two black holes with charges $\ p^{1}=\ p^{2}=\ p^{3}=\ q_{0}=1$%
, and $\ p^{1}=\ p^{2}=1$, $\ p^{3}=-1$, $\ q_{0}=-1$, respectively, their $|\psi
^{\prime \prime }\rangle $ are both $\sqrt{2}(|000\rangle +|111\rangle )$.

\subsubsection{Two BPS black holes with $p^{1}p^{2}p^{3}q_{0}>0$ differing
only by signs are LU equivalent}

For $H_{i}$ and $H_{i}^{\prime }$, $i=1,2,3,4$, we will show that $|\psi
\rangle $ in Eq. \eqref{ghz-1} is transformed into the same $|\psi ^{\prime
\prime }\rangle $ under LU. Thus, $|\psi \rangle $ in Eq. \eqref{ghz-1} with 
$p^{1}p^{2}p^{3}q_{0}>0$ are LU equivalent \cite{dli-jpa}.\textbf{\ }The
following is the argument.

Clearly, $|\rho _{j}|$, $j=0,1,2,3,4$, are independent of the sign of the
charges. Therefore, $|\rho _{j}|$ are the same for $H_{i}$ and $%
H_{i}^{\prime }$, $i=1,2,3,4$.

Similarly, we can show $e^{i\varphi }$ is the same for any $H_{i}$ and $%
H_{i}^{\prime }$. From Eq. \eqref{ru-1}, we know that $\omega _{1}=\pm \pi
/2$, and\ $\omega _{1}$ is independent of the signs of the charges.
Moreover, for each of the four cases A), B), C), and D), $e^{i\varphi}$ is
the same for $H_{i}$ and $H_{i}^{\prime }$, $i=1,2,3,4$.

\subsection{A non-BPS black hole with $p^{1}p^{2}p^{3}q_{0}<0$ and\ a BPS
black hole with $p^{1}p^{2}p^{3}q_{0}>0$ differing only by signs are LU
equivalent}

For example, for one state $|\psi_1 \rangle $ with $p^{1}p^{2}p^{3}q_{0}<0$
(e.g., $p^{1}=p^{2}=p^{3}=1$ and $ q_{0}=-4$), and another state $|\psi_2 \rangle $ with $p^{1}p^{2}p^{3}q_{0}>0$ (e.g., $p^{1}=
p^{2}= p^{3}=1$ and $ q_{0}=4$), we can show that they are LU equivalent.

Let $ |\psi'\rangle= U^{A}\otimes U^{B}\otimes U^{C}|\psi_1 \rangle$, and $|\psi''\rangle = W^{A}\otimes W^{B}\otimes W^{C}|\psi_2 \rangle$.
We first show that $|\psi ^{\prime }\rangle =|\psi ^{\prime\prime }\rangle $.

It is easy to see $|\eta _{j}|=|\rho _{j}|$, $j=0,1,2,3,4$. We next
show $e^{i\varphi }$ in Eq. \eqref{ASD-2-1} and $e^{i\phi }$ in Eq. %
\eqref{ASD-1-} are equal. From eqs. \eqref{ada-1} and \eqref{ru-1}, we know $%
\theta _{1}=0$ and $\omega _{1}=-\pi /2$ whenever $%
(p^{1})^{2}+(p^{2})^{2}-(p^{3})^{2}-q_{0}^{2}>0$, while $\theta _{1}=\pi $
and $\omega _{1}=\pi /2$ whenever $%
(p^{1})^{2}+(p^{2})^{2}-(p^{3})^{2}-q_{0}^{2}<0$. Then, one can see that $%
e^{i\varphi }=e^{i\phi }$ for each of the four cases A), B), C), and D).
Therefore, $|\psi ^{\prime }\rangle =|\psi ^{\prime \prime }\rangle $.

\vspace{5pt}

Then from $U^{A}\otimes U^{B}\otimes U^{C}|\psi_1 \rangle =W^{A}\otimes W^{B}\otimes W^{C}|\psi_2 \rangle$, we can derive that $$|\psi_2 \rangle
=(W^{A})^{-1}U^{A}\otimes (W^{B})^{-1}U^{B}\otimes
(W^{C})^{-1}U^{C}|\psi_1 \rangle $$

$(W^{A})^{-1}U^{A}$ is still a unitary operator, and so is $(W^{B})^{-1}U^{B}$ and $(W^{C})^{-1}U^{C}$. Thus, $|\psi_1 \rangle $  and $|\psi_2 \rangle $ are LU equivalent.

\vspace{10pt}

\textit{\bf Proposition 1}. {\it (Sufficient condition for LU\ equivalence of two
black holes (BPS or non-BPS) with $p^{1}p^{2}p^{3}q_{0}\neq 0$) } Between two black holes (BPS or non-BPS) with $p^{1}p^{2}p^{3}q_{0}\neq 0$,
if their only difference is in the signs of the charges, then they are LU
equivalent.
\vspace{5pt}

{\bf Remark 1.} BPS black holes (with $p^{1}p^{2}p^{3}q_{0}>0$) and
non-BPS black holes (with $p^{1}p^{2}p^{3}q_{0}<0$) are regarded as two inequivalent
subclasses of GHZ SLOCC class \cite{Linde, Borsten09}. Note that the two subclasses of GHZ SLOCC\ class are not classified under LU. Proposition 1 shows that they could be LU equivalent. 

\vspace{5pt}

{\bf Remark 2.} Via Eq. (44) of \cite{Behrndt} and Eq. (43) of \cite{Levay}, one
can see that any two LU equivalent black holes with charges $%
p^{1},p^{2},p^{3},$\ and $q_{0}$\ have the same entanglement, black hole
entropy, area of the black hole horizon, and black hole mass.\textbf{\ }

\section{Classify \textquotedblleft large\textquotedblright\ black holes into seven families}
\label{sec:7families}

Although the physical basis for the black hole--qubit correspondence is not completely understood (\cite{Levay10}),
it is widely accepted that additional insight into the string black hole theory or quantum information theory can be obtained by
exploiting methods and techniques of the other (\cite{Duff-Fe,Duff,
Levay07-prd-75, Duff-Ferrara, Levay07, Bellucci07, Duff08, Borsten-Duff,
Borsten-Duff-prd, Bellucci, Borsten09}), and a more
detailed classification than the SLOCC classification of three qubits is required \cite
{Linde}. In this section, by exploiting the LU classification of three
qubits we classify black holes with non-vanishing charges $
p^{1},p^{2},p^{3}, $ and $q_{0}$. Because $|\psi ^{\prime }\rangle =|\psi
^{\prime \prime }\rangle $, we only need to classify the $|\psi ^{\prime
}\rangle $ in Eq. \eqref{ASD-1-}. Then, we classify \textquotedblleft
large\textquotedblright\ black holes into seven families that are pairwise LU
inequivalent. {The detailed derivation of the seven families can be found in Appendix \ref{sec:proof}.}

\textit{Family 1}. Let Family 1 consist of the black holes satisfying the
following equation: 
\begin{equation}
(p^{1})^{2}=(p^{2})^{2}=(p^{3})^{2}=q_{0}^{2}.  \label{cond-1}
\end{equation}

Note that Eq. \eqref{cond-1} holds if and only if $|p^{1}p^{3}|=|p^{2}q_{0}|$%
, $|p^{2}p^{3}|=|p^{1}q_{0}|$ and $%
(p^{1})^{2}+(p^{2})^{2}-(p^{3})^{2}-q_{0}^{2}=0$. In other words, Eq. 
\eqref
{cond-1} holds if and only if $\eta _{1}=\eta _{2}=\eta _{3}=0$. Thus, Family 1 corresponds to the states of the following form:
\begin{equation}
|p^{1}|\sqrt{2}(|000\rangle +|111\rangle ).  \label{coro-1}
\end{equation}

Generally speaking, $|\psi \rangle $ in Eq. \eqref{ghz-1} cannot be
transformed into the canonical GHZ state under LU. $|\psi \rangle $ in Eq. %
\eqref{ghz-1} can be transformed into the canonical GHZ state under LU if
and only if Eq. \eqref
{cond-1} holds, under which the canonical GHZ state is $|p^{1}|\sqrt{2}
(|000\rangle +|111\rangle )$. Specially, when $%
(p^{1})^{2}=(p^{2})^{2}=(p^{3})^{2}=(q_{0})^{2}=1$, the canonical GHZ is $%
\sqrt{2}(|000\rangle +|111\rangle )$.

\textit{Family 2}. Let Family 2 consist of the black holes satisfying the
following equation: 
\begin{equation}
(p^{1})^{2}=(p^{3})^{2},(p^{2})^{2}=q_{0}^{2},(p^{1})^{2}\neq (p^{2})^{2}.
\label{cond-2}
\end{equation}%
Note that Eq. \eqref{cond-2} holds if and only if $%
(p^{1})^{2}+(p^{2})^{2}-(p^{3})^{2}-q_{0}^{2}=0$ (i.e., $\eta _{1}=0$), $%
|p^{2}p^{3}|=|p^{1}q_{0}|$ (i.e., $\eta _{3}=0$), but $|p^{1}p^{3}|\neq
|p^{2}q_{0}|$ (i.e., $\eta _{2}\neq 0$). In other words, Eq. \eqref{cond-2}
holds if and only if $\eta _{1}=\eta _{3}=0$, but $\eta _{2}\neq 0$. Thus,
Family 2 corresponds to the states of the following form:
\begin{equation}
\eta _{0}|000\rangle +|\eta _{2}||101\rangle +|\eta _{4}||111\rangle .
\label{coro-2}
\end{equation}

\textit{Family 3}. Let Family 3 consist of the black holes satisfying the
following equation: 
\begin{equation}
(p^{2})^{2}=(p^{3})^{2},(p^{1})^{2}=q_{0}^{2},(p^{1})^{2}\neq (p^{2})^{2}.
\label{cond-3}
\end{equation}
Note that Eq. \eqref{cond-3} holds if and only if $%
(p^{1})^{2}+(p^{2})^{2}-(p^{3})^{2}-q_{0}^{2}=0$ (i.e., $\eta _{1}=0$), $%
|p^{1}p^{3}|=|p^{2}q_{0}|$ (i.e., $\eta _{2}=0$), but $|p^{2}p^{3}|\neq
|p^{1}q_{0}|$ (i.e., $\eta _{3}\neq 0$). In other words, Eq. (\ref{cond-3})
holds if and only if $\eta _{1}=\eta _{2}=0$, but $\eta _{3}\neq 0$. Thus,
Family 3 corresponds to the states of the following form:
\begin{equation}
\eta _{0}|000\rangle +|\eta _{3}||110\rangle +|\eta _{4}||111\rangle .
\label{coro-3}
\end{equation}

\textit{Family 4}. Let Family 4 consist of the black holes satisfying the
following equation: 
\begin{equation}
(p^{1})^{2}=(p^{2})^{2},(p^{3})^{2}=q_{0}^{2},(p^{1})^{2}\neq (p^{3})^{2}
\label{cond-4}
\end{equation}
Note that Eq. \eqref{cond-4} holds if and only if $%
(p^{1})^{2}+(p^{2})^{2}-(p^{3})^{2}-q_{0}^{2}\neq 0$, but $%
|p^{2}p^{3}|=|p^{1}q_{0}|$, and $|p^{1}p^{3}|=|p^{2}q_{0}|$. In other words,
Eq. \eqref{cond-4} holds if and only if $\eta _{2}=\eta _{3}=0$, but $\eta
_{1}\neq 0$. Thus, Family 4 corresponds to the states of the following form:
\begin{equation}
\eta _{0}|000\rangle \pm |\eta _{1}||100\rangle +|\eta _{4}||111\rangle.
\label{coro-4}
\end{equation}

\textit{Family 5}. Let Family 5 consist of the black holes satisfying the
following equation: 
\begin{equation}
(p^{2})^{2}\neq (p^{3})^{2},(p^{3})^{2}\neq
q_{0}^{2},|p^{1}p^{3}|=|p^{2}q_{0}|  \label{cond-5}
\end{equation}
Note that Eq. \eqref{cond-5} holds if and only if $%
(p^{1})^{2}+(p^{2})^{2}-(p^{3})^{2}-q_{0}^{2}\neq 0,|p^{2}p^{3}|\neq
|p^{1}q_{0}|,|p^{1}p^{3}|=|p^{2}q_{0}|$. In other words, Eq. \eqref{cond-5}
holds if and only if $\eta _{1}\neq 0$, $\eta _{2}=0$, $\eta _{3}\neq 0$.
For example, $(p^{1})^{2}=4,(p^{2})^{2}=1,(p^{3})^{2}=16,q_{0}^{2}=64$,
which satisfies Eq. \eqref{cond-5}. Thus, Family 5 corresponds to the states of the following form:

\begin{equation}
\eta _{0}|000\rangle \pm |\eta _{1}||100\rangle +|\eta _{3}||110\rangle
+|\eta _{4}||111\rangle .  \label{coro-5}
\end{equation}

\textit{Family 6}. Let Family 6 consist of the black holes satisfying the
following equation: 
\begin{equation}
(p^{2})^{2}\neq q_{0}^{2},(p^{3})^{2}\neq
q_{0}^{2},|p^{2}p^{3}|=|p^{1}q_{0}|.  \label{cond-6}
\end{equation}
Note that Eq. \eqref{cond-6} holds if and only if $%
(p^{1})^{2}+(p^{2})^{2}-(p^{3})^{2}-q_{0}^{2}\neq
0,|p^{2}p^{3}|=|p^{1}q_{0}|,|p^{1}p^{3}|\neq |p^{2}q_{0}|$. In other words,
Eq. \eqref{cond-6} holds if and only if $\eta _{1}\neq 0$, $\eta _{2}\neq 0$
, $\eta _{3}=0$. Thus, Family 6 corresponds to the states of
the following form:

\begin{equation}
\eta _{0}|000\rangle \pm |\eta _{1}||100\rangle +|\eta _{2}||101\rangle
+|\eta _{4}||111\rangle .  \label{coro-6}
\end{equation}

\textit{Family 7}. Let Family 7 consist of the black holes satisfying the
following equation: 
\begin{equation}
|p^{1}p^{3}|\neq |p^{2}q_{0}|,|p^{2}p^{3}|\neq |p^{1}q_{0}|.  \label{cond-7}
\end{equation}%
Note that Eq. \eqref{cond-7} holds if and only if $\eta _{2}\eta _{3}\neq 0$
but $\eta _{1}$ may vanish. Thus, Family 7 corresponds to the states of
the following form:
\begin{equation}
\eta _{0}|000\rangle +|\eta _{1}|e^{i\phi }|100\rangle +|\eta
_{2}||101\rangle +|\eta _{3}||110\rangle +|\eta _{4}||111\rangle ,
\label{coro-7}
\end{equation}%
where $\eta _{1}$ may vanish. 

\ignore{
To see why $\eta _{1}$ may vanish, we consider this example: let

\begin{eqnarray}
|\varsigma \rangle &=&(1/2)(|000\rangle +|101\rangle +|110\rangle
+|111\rangle ), \\
|\varsigma ^{\prime }\rangle &=&\frac{\sqrt{2}}{2}|000\rangle -\frac{\sqrt{2}%
}{4}|100\rangle +\frac{\sqrt{2}}{4}|101\rangle  \notag \\
&&+\frac{\sqrt{2}}{4}|110\rangle +\frac{\sqrt{2}}{4}|111\rangle .
\end{eqnarray}%

We can show 
\begin{equation}
|\varsigma ^{\prime }\rangle =H\otimes H\otimes H|\varsigma \rangle ,
\end{equation}%
where $H$ is the Hadamard matrix. Therefore, $|\varsigma ^{\prime }\rangle $
and $|\varsigma \rangle $ are LU equivalent. We can see that for $|\varsigma
\rangle $, $\eta _{1}=0$, whereas for $|\varsigma ^{\prime }\rangle $, $\eta
_{1}\neq 0$. 
}

\vspace{10pt}

The LU invariants $J_{1}=|\lambda_{1}\lambda _{4}e^{i\varphi }-\lambda _{2}\lambda _{3}|^{2}$, $
J_{i}=(\lambda _{0}\lambda _{i})^{2}$, $i=2,3$, were defined in \cite{Acin00}. In Table \ref{tbl:LU}, we calculate $
J_{i}$ for states in eqs. \eqref{coro-1}, \eqref{coro-2}, \eqref{coro-3}, \eqref
{coro-4}, \eqref{coro-5}, \eqref{coro-6}, and \eqref{coro-7}. From Table \ref{tbl:LU}, we obtain the following proposition:

\textbf{Proposition 2.} Let Group 1 (resp. 2, 3, 4, 5, 6, 7) consist of the
states in eqs. \eqref{coro-1} (resp. \eqref{coro-2}, \eqref{coro-3}, %
\eqref{coro-4}, \eqref{coro-5}, \eqref{coro-6}, \eqref{coro-7}). Then, the
seven groups are pairwise LU inequivalent.

\begin{table}[tbp]
\caption{LU invariants $J_1$, $J_2$ and $J_3$. $G_{i}$ stands for Group $i$, corresponding to $F_i$ in table \ref{tbl:family}, for $i=1, \ldots, 7$. The ``Don't care" term means it can be either $=0$ or $\ne0$.}
\label{tbl:LU}%
\begin{tabular}{|l|c|c|c|c|c|c|c|}
\toprule & $G_{1}$ & $G_{2}$ & $G_{3}$ & $G_{4}$ & $G_{5}$ & $G_{6}$ & $
G_{7} $ \\ 
\midrule $J_{1}$ & 0 & 0 & 0 & $\neq 0$ & $\neq 0$ & $\neq 0$ & Don't care \\ 
\midrule $J_{2}$ & 0 & $\neq 0$ & 0 & 0 & 0 & $\neq 0$ & $\neq 0$ \\ 
\midrule $J_{3}$ & 0 & 0 & $\neq 0$ & 0 & $\neq 0$ & 0 & $\neq 0$ \\ 
\bottomrule 
\end{tabular}
\end{table}

\vspace{5pt}
From Proposition 2, we obtain the following proposition:

\vspace{0.1in}

\textbf{Proposition 3.} The black holes with non-vanishing charges $q_{0}$, $%
p^{1}$, $p^{2}$, $p^{3}$ are classified into seven inequivalent families
(see Table \ref{tbl:family}).

\begin{table}[tbp]
\caption{Criteria for the seven families of black holes. $F_{i}$ stands for Family $i$.}
\label{tbl:family}
\begin{tabular}{|l|l|}
\toprule
& Criteria \\ 
\midrule $F_{1}$ & $(p^{1})^{2}=(p^{2})^{2}=(p^{3})^{2}=q_{0}^{2}$ \\ 
\midrule $F_{2}$ & $
(p^{1})^{2}=(p^{3})^{2},(p^{2})^{2}=q_{0}^{2},(p^{1})^{2}\neq (p^{2})^{2}$
\\ 
\midrule $F_{3}$ & $
(p^{2})^{2}=(p^{3})^{2},(p^{1})^{2}=q_{0}^{2},(p^{1})^{2}\neq (p^{2})^{2}$
\\ 
\midrule $F_{4}$ & $
(p^{1})^{2}=(p^{2})^{2},(p^{3})^{2}=q_{0}^{2},(p^{1})^{2}\neq (p^{3})^{2}$
\\ 
\midrule $F_{5}$ & $(p^{2})^{2}\neq
(p^{3})^{2},(p^{3})^{2}\neq q_{0}^{2},\frac{|p^{1}|}{|q_{0}|}=\frac{|p^{2}|}{
|p^{3}|}$ \\ 
\midrule $F_{6}$ & $(p^{2})^{2}\neq q_{0}^{2}, (p^{3})^{2}\neq q_{0}^{2},\frac{|p^{2}|}{|q_{0}|}=\frac{|p^{1}|}{
|p^{3}|}$ \\ 
\midrule $F_{7}$ & $\frac{|p^{1}|}{|q_{0}|}\neq \frac{|p^{2}|}{|p^{3}|},
\frac{|p^{2}|}{|q_{0}|}\neq \frac{|p^{1}|}{|p^{3}|}$ \\ 
\bottomrule  
\end{tabular}
\end{table}

\subsection{Some properties for LU classification of the black holes with
charges $q_{0}$, $p^{1}$, $p^{2}$, and $p^{3}$}

It is known that there are infinite LU equivalence classes for the black holes
with charges $q_{0}$, $p^{1}$, $p^{2}$, and $p^{3}$. Via LU transformation,
the infinite LU classes are classified into seven families in this section.
Thus, some families include infinite LU equivalence classes. 

\vspace{5pt}

(i). Clearly, two LU equivalent black holes must be in the same family while
two black holes belonging to different families must be LU inequivalent, i.e.,

\vspace{5pt}

\begin{itemize}
\item[] 
\begin{itemize}
\item Two black holes are LU equivalent  $\overset{\longrightarrow}{\nleftarrow}$ they belong to the same family. Note that two black holes being in the same family does not necessarily imply that they are LU equivalent.
\item Two black holes are from different families $\rightarrow$ they belong to different LU equivalence classes (i.e., they are LU inequivalent).
\end{itemize}
\end{itemize}

\vspace{5pt}
(ii). Via Eq. (44) of \cite{Behrndt} and Eq. (43) of \cite{Levay}, one can
see that for two LU equivalent black holes with charges $q_{0}$, $p^{1}$, $%
p^{2}$, $p^{3}$, they have the same entanglement, the same black hole
entropy, the same area of the black hole horizon, and the same black hole
mass.  

\vspace{5pt}
(iii). For Family 1, we show Family 1 has the maximal entanglement, entropy
of black holes, area of the black hole horizon, and black hole mass for
normalised black holes

\textbf{Remark 3.} In \cite{Linde}, Kallosh and Linde showed that the supergravity charges $q_{i}$ and $p^i$ $(i=0,1,2,3)$
originate from the number of D0, D2, D4 and D6 branes: the number nD0 of D0 branes is $q_{0}$, and the numbers kD2, mD2, lD2 of D2 branes are $q_{1}$, $q_{2}$, and $q_{3}$, respectively; the number nD6 of D6 branes is $p^{0}$, and the numbers kD4, mD4, lD4 of D4 branes are $p^{1}$, $p^{2}$, and $p^{3}$, respectively. Therefore, Eqs. (\ref{cond-1}, \ref{cond-2}, \ref{cond-3}, \ref
{cond-4}, \ref{cond-5}, \ref{cond-6}, \ref{cond-7}) can be rewritten by
replacing $p^{1}$, $p^{2}$, $p^{3}$, $q_{0}$ with kD4, mD4, lD4, nD0,
respectively. For example, Eq. \eqref{cond-1} can be written as $
(k_{D4})^{2}=(m_{D4})^{2}=(l_{D4})^{2}=(n_{D0})^{2}$, etc.

\section{A relation between two entropies: von Neumann entanglement entropy
in quantum information theory and the entropy of black holes in string
theory}
\label{sec:2entropy}

The von Neumann entanglement entropy is an important entanglement measure of
the degree of entanglement between two subsystems. In \cite{Levay, Duff}, the
authors indicated that since the near horizon geometry of black holes is AdS$
_{2} \times S^{2}$, using the idea of AdS/CFT (anti-de Sitter/conformal field theory) holography, one might expect a relation between entanglement entropy in quantum information theory and black hole entropy in string theory. However, until this day, the
correspondence between these two different physical notions, black hole
entropy and von Neumann entanglement entropy, is not well-understood \cite{Levay, Duff}.

It is known that Cayley's hyperdeterminant provides an interesting mathematical
connection between the entropy of black holes in string theory and the quantum entanglement measure 3-tangle \cite{Duff}. However, in any event, the 3-tangle is a different notion from entanglement entropy for three qubits. Thus, ``The appearance
of the Cayley hyperdeterminant in these two different contexts of stringy black hole entropy and 3-tangle for three qubits remains, for the moment, a purely mathematical coincidence" \cite{Duff}.

In the following, we propose a correspondence between the black hole entropy and von Neumann entanglement entropy.

In \cite{dli-qip}, the following von Neumann entanglement entropy of three qubits $A$, $B$, and $C$ is given: 

\begin{eqnarray}
S(\rho _{\nu })    &=& -(\gamma _{\nu }^{(1)}\ln \gamma _{\nu }^{(1)}+\gamma _{\nu
}^{(2)}\ln \gamma _{\nu }^{(2)}),
\end{eqnarray}
where $\nu \in \{A,B,C\}$, $\gamma _{\nu }^{(1)} = \frac{1+\sqrt{1-4\alpha _{\nu }}}{2}$, and $\gamma _{\nu }^{(2)} = \frac{1-\sqrt{1-4\alpha _{\nu }}}{2}$, in which $0 \leq \alpha _{\nu }\leq 1/4$ can be calculated as follows,

\begin{eqnarray}
\alpha _{A} &=&\frac{\tau _{AB}+\tau _{AC}+\tau _{ABC}}{4}=\frac{\tau
_{A(BC)}}{4}, \\
\alpha _{B} &=&\frac{\tau _{AB}+\tau _{BC}+\tau _{ABC}}{4}=\frac{\tau
_{B(AC)}}{4}, \\
\alpha _{C} &=&=\frac{\tau _{AC}+\tau _{BC}+\tau _{ABC}}{4}=\frac{\tau
_{C(AB)}}{4}.
\end{eqnarray}%

\vspace{5pt}

Via the classical supergravity entropy formula \cite{Linde}, we obtain 
\begin{equation}
\tau _{ABC}=\frac{4S_{bh}^{2}}{\pi ^{2}}.
\end{equation}

Thus,\ from the above equations we obtain a relation between the von Neumann
entanglement entropy $S(\rho _{\nu })$ and the black hole entropy $S_{bh}$. We have the following approximate relations:

\begin{eqnarray*}
2S(\rho _{A}) &\approx &2\ln 2-1+\tau _{AB}+\tau _{AC}+\frac{4S_{bh}^{2}}{%
\pi ^{2}}, \\
2S(\rho _{B}) &\approx &2\ln 2-1+\tau _{AB}+\tau _{BC}+\frac{4S_{bh}^{2}}{%
\pi ^{2}}, \\
2S(\rho _{C}) &\approx &2\ln 2-1+\tau _{AC}+\tau _{BC}+\frac{4S_{bh}^{2}}{%
\pi ^{2}}.
\end{eqnarray*}

Let us consider the average 2-tangles and the average von Neumann
entanglement entropy

\begin{eqnarray}
A &=&\frac{\tau _{AB}+\tau _{AC}+\tau _{BC}}{3},  \label{average-1} \\
m_{vn} &=&\frac{S(\rho _{A})+S(\rho _{B})+S(\rho _{C})}{3}.
\end{eqnarray}

Thus, we obtain a relation between the average entanglement entropy (i.e., average von Neumann's entanglement entropy) and the entropy of black holes:

\begin{equation}
m_{vn}\approx (\ln 2-\frac{1}{2})+A+\frac{2S_{bh}^{2}}{\pi ^{2}}
\label{relat-0}
\end{equation}

{\bf Discussion:}
\begin{itemize}
\item Since the W SLOCC state has the maximal average 2-tangles $A=\frac{3\ln 3-2\ln 2}{3}%
=\allowbreak 0.636\,51$, one can see from Eq. \eqref{relat-0} that the
difference between the average von Neumann entanglement entropy and $\frac{
2S_{bh}^{2}}{\pi ^{2}}$ reaches the maximal $\allowbreak 0.829\,66$.

\vspace{5pt}

\item Note that $\ln 2-1/2=\allowbreak 0.193\,15$. Ignoring $\ln 2-1/2$ in Eq. \eqref{relat-0}, we obtain 
\begin{equation}
m_{vn}-A\approx \frac{2S_{bh}^{2}}{\pi ^{2}}.  \label{relat-1}
\end{equation}

For the general GHZ states $\lambda _{0}|000\rangle +\lambda _{4}|111\rangle 
$, we have $A=0$, thus, 
\begin{equation}
m_{vn}\approx \frac{2S_{bh}^{2}}{\pi ^{2}}
\end{equation}
\end{itemize}

\section{Black holes with the maximal von Neumann entanglement entropy}
\label{sec:maxvon}

Von Neumann entropy has direct impact on our understanding black holes and
is an important mean to describe black holes. For example, von Neumann
entropy of black holes was computed \cite{geo}. In this section, we derive
the black holes with the maximal von Neumann entanglement entropy $S(\rho
_{A})=S(\rho _{B})=S(\rho _{C})=\ln 2$, the maximal von Neumann entanglement
entropy $S(\rho _{x})=S(\rho _{y})=\ln 2$ while $S(\rho _{z})<\ln 2$, and
the maximal von Neumann entanglement entropy $S(\rho _{x})=\ln 2$, while $
S(\rho _{y})<\ln 2$ and $S(\rho _{z})<\ln 2$, where $xyz=\{ABC,BCA,CAB\}$.
Thus, we study the properties of von Neumann entanglement entropy for each
of seven families. For example, we show that Family 1 includes all the black
holes which have the maximal von Neumann entanglement entropy $S(\rho
_{A})=S(\rho _{B})=S(\rho _{C})=\ln 2$.

As stated before, the black hole with four non-vanishing charges $
q_{0}$, $p^{1}$, $p^{2}$, and $p^{3}$ corresponds to a three-qubit pure
state, 
\begin{equation}
|\psi \rangle =-p^{1}|001\rangle -p^{2}|010\rangle -p^{3}|100\rangle
+q_{0}|111\rangle .
\end{equation}
The $|\psi \rangle $ belongs to GHZ SLOCC\ class. For $p^{1}p^{2}p^{3}q_{0}<0
$, we list the above $|\psi ^{\prime }\rangle $ as follows.

\begin{equation}
|\psi ^{\prime }\rangle =\eta _{0}|000\rangle \pm |\eta _{1}||100\rangle
+|\eta _{2}||101\rangle +|\eta _{3}||110\rangle +|\eta _{4}||111\rangle .
\end{equation}

\smallskip

From Proposition 4 in Appendix D, we know $\eta _{0}^{2}+\eta _{1}^{2}+\eta
_{2}^{2}+\eta _{3}^{2}+\eta
_{4}^{2}=(p_{1})^{2}+(p_{2})^{2}+(p_{3})^{2}+q_{0}^{2}$. Therefore, the
following $\mu |\psi ^{\prime }\rangle $ is SD of $\mu |\psi \rangle $. 
\begin{equation}
\mu |\psi ^{\prime }\rangle =\mu \eta _{0}|000\rangle \pm \mu |\eta
_{1}||100\rangle +\mu |\eta _{2}||101\rangle +\mu |\eta _{3}||110\rangle
+\mu |\eta _{4}||111\rangle .
\end{equation}

\textit{Family 1. }Family 1 corresponds to the states of the following form: 
\begin{equation}
|\nu \rangle =|p^{1}|\sqrt{2}(|000\rangle +|111\rangle ).  \label{vne-coro-1}
\end{equation}

Under the condition in Eq. (\ref{cond-1}) $\mu |\nu \rangle $ is normal and

\begin{equation}
\mu |\nu \rangle =|\text{GHZ}\rangle =\frac{1}{\sqrt{2}}(|000\rangle
+|111\rangle ).
\end{equation}

By Lemma 1 in Appendix C, $\mu |\nu \rangle $ (i.e. $|$GHZ$\rangle $) has
the maximal von Neumann entanglement entropy $S(\rho _{A})=S(\rho
_{B})=S(\rho _{C})=\ln 2$. Therefore, under the factor $\mu $ each black
hole of Family 1 has the maximal von Neumann entanglement entropy $S(\rho
_{A})=S(\rho _{B})=S(\rho _{C})=\ln 2$.

\textit{Family 2}. Family 2 corresponds to the states of the following form: 
\begin{equation}
|\varsigma \rangle =\eta _{0}|000\rangle +|\eta _{2}||101\rangle +|\eta
_{4}||111\rangle .
\end{equation}

Clearly, $\mu |\varsigma \rangle $ is normal and 
\begin{equation}
\mu |\varsigma \rangle =\mu \eta _{0}|000\rangle +\mu |\eta _{2}||101\rangle
+\mu |\eta _{4}||111\rangle .
\end{equation}
\smallskip
By Proposition 5 in Appendix D and the condition $%
(p^{1})^{2}+(p^{2})^{2}-(p^{3})^{2}-q_{0}^{2}=0$ for Family 2, obtain $\mu
\eta _{0}=\frac{1}{\sqrt{2}}$. Thus, 
\begin{equation}
\mu |\varsigma \rangle =\frac{1}{\sqrt{2}}|000\rangle +\mu |\eta
_{2}||101\rangle +\mu |\eta _{4}||111\rangle .
\end{equation}

By Corollary 1 in Appendix C, $\mu |\varsigma \rangle $ has the maximal von
Neumann entanglement entropy $S(\rho _{A})=S(\rho _{C})=\ln 2$ while $S(\rho
_{B})<\ln 2$. Thus, for Family 2, under the factor $\mu $ each black hole
has the maximal von Neumann entanglement entropy $S(\rho _{A})=S(\rho
_{C})=\ln 2$ while $S(\rho _{B})<\ln 2$.

\textit{Family 3}. Family 3 corresponds to the states of the following form: 
\begin{equation}
|\iota \rangle =\eta _{0}|000\rangle +|\eta _{3}||110\rangle +|\eta
_{4}||111\rangle .
\end{equation}

Clearly, $\mu |\iota \rangle $ is normal and 
\begin{equation}
\mu |\iota \rangle =\mu \eta _{0}|000\rangle +\mu |\eta _{3}||110\rangle
+\mu |\eta _{4}||111\rangle .
\end{equation}

By Proposition 5 in Appendix D and the condition $%
(p^{1})^{2}+(p^{2})^{2}-(p^{3})^{2}-q_{0}^{2}=0$ for Family 3, obtain $\mu
\eta _{0}=\frac{1}{\sqrt{2}}$. Thus, 
\begin{equation}
\mu |\iota \rangle =\frac{1}{\sqrt{2}}|000\rangle +\mu |\eta
_{3}||110\rangle +\mu |\eta _{4}||111\rangle .
\end{equation}

By Corollary 2 in Appendix C, $\mu |\iota \rangle $ has the maximal von
Neumann entanglement entropy $S(\rho _{A})=S(\rho _{B})=\ln 2$ while $S(\rho
_{C})<\ln 2$. Thus, for Family 3, under the factor $\mu $ each black hole
has the maximal von Neumann entanglement entropy $S(\rho _{A})=S(\rho
_{B})=\ln 2$ while $S(\rho _{C})<\ln 2$.

\textit{Family 4}. Family 4 corresponds to the states of the following form: 
\begin{equation}
|\vartheta \rangle =\eta _{0}|000\rangle \pm |\eta _{1}||100\rangle +|\eta
_{4}||111\rangle .  \label{vne-coro-4}
\end{equation}

Clearly, $\mu |\vartheta \rangle $ is normal and 
\begin{equation}
\mu |\vartheta \rangle =\mu \eta _{0}|000\rangle \pm \mu |\eta
_{1}||100\rangle +\mu |\eta _{4}||111\rangle .
\end{equation}

From $(p^{1})^{2}=(p^{2})^{2},(p^{3})^{2}=q_{0}^{2},$ and $(p^{1})^{2}\neq
(p^{3})^{2}$ for Family 4, a calculation yields that $\mu |\eta _{4}|=\frac{1%
}{\sqrt{2}}$. Then, 
\begin{equation}
\mu |\vartheta \rangle =\mu \eta _{0}|000\rangle \pm \mu |\eta
_{1}||100\rangle +\frac{1}{\sqrt{2}}|111\rangle .
\end{equation}

By Corollary 4 in Appendix C, $\mu |\vartheta \rangle $ has the maximal von
Neumann entanglement entropy $S(\rho _{B})=S(\rho _{C})=\ln 2$ while $S(\rho
_{A})<\ln 2$. Thus, for Family 4 under the factor $\mu $ each black hole has
the maximal von Neumann entanglement entropy $S(\rho _{B})=S(\rho _{C})=\ln
2 $ while $S(\rho _{A})<\ln 2$.

\textit{Family 5}. Family 5 corresponds to the states of the following form:

\begin{equation}
|\varkappa \rangle =\eta _{0}|000\rangle \pm |\eta _{1}||100\rangle +|\eta
_{3}||110\rangle +|\eta _{4}||111\rangle .
\end{equation}

Clearly, $\mu |\varkappa \rangle $ is normal. By Lemmas 1-8 in Appendix C, $%
\mu |\varkappa \rangle $ has the von Neumann entanglement entropy $S(\rho
_{y})<\ln 2$, $y=A,B,C$, and under the factor $\mu $ each black hole of
Family 5 also has the von Neumann entanglement entropy $S(\rho _{y})<\ln 2$, 
$y=A,B,C$.

\textit{Family 6}. Family 6 corresponds to the states of the following form:

\begin{equation}
|R\rangle =\eta _{0}|000\rangle \pm |\eta _{1}||100\rangle +|\eta
_{2}||101\rangle +|\eta _{4}||111\rangle .  \label{vne-coro-6}
\end{equation}

Clearly, $\mu |R\rangle $ is normal. By Lemmas 1-8 in Appendix C, $\mu
|R\rangle $ has the von Neumann entanglement entropy $S(\rho _{y})<\ln 2$, $%
y=A,B,C$ and under the factor $\mu $ each black hole of Family 6 also has
the von Neumann entanglement entropy $S(\rho _{y})<\ln 2$, $y=A,B,C$.

\textit{Family 7}. Family 7 corresponds to the states of the following form: 
\begin{equation}
|T\rangle =\eta _{0}|000\rangle \pm |\eta _{1}||100\rangle +|\eta
_{2}||101\rangle +|\eta _{3}||110\rangle +|\eta _{4}||111\rangle ,
\label{vne-coro-7}
\end{equation}

Then, $\mu |T\rangle $ is normal and 
\begin{equation}
\mu |T\rangle =\mu \eta _{0}|000\rangle \pm \mu |\eta _{1}||100\rangle +\mu
|\eta _{2}||101\rangle +\mu |\eta _{3}||110\rangle +\mu |\eta
_{4}||111\rangle ,
\end{equation}

Let us divide Family 7 into four subfamilies.

Family 7.1. Let Family 7.1 consist of the black holes satisfying the
following equation: 
\begin{equation}
(p^{1})^{2}+(p^{2})^{2}-(p^{3})^{2}-q_{0}^{2}=0.
\end{equation}

Note that when $(p^{1})^{2}+(p^{2})^{2}-(p^{3})^{2}-q_{0}^{2}=0$, then $%
|\eta _{1}|=0$. Thus, Family 7.1 corresponds to the states of the following
form

\begin{equation}
|T_{1}\rangle =\eta _{0}|000\rangle +|\eta _{2}||101\rangle +|\eta
_{3}||110\rangle +|\eta _{4}||111\rangle .
\end{equation}

One can see that $\mu |T_{1}\rangle $ is normal and $\mu \eta _{0}=\frac{1}{%
\sqrt{2}}$ by Proposition 5 in Appendix D. Thus,

\begin{equation}
\mu |T_{1}\rangle =\frac{1}{\sqrt{2}}|000\rangle +\mu |\eta _{2}||101\rangle
+\mu |\eta _{3}||110\rangle +\mu |\eta _{4}||111\rangle ,  \label{Fam-7-1}
\end{equation}

Clearly, $|\mu T_{1}\rangle $ has the maximal von Neumann entanglement
entropy $S(\rho _{A})(=\ln 2)$ while $S(\rho _{B})<\ln 2$ and $S(\rho
_{C})<\ln 2$ by Corollary 3 in Appendix C. Therefore, under the factor $\mu $%
\ each black hole of Family 7.1 has the maximal von Neumann entanglement
entropy $S(\rho _{A})(=\ln 2)$ while $S(\rho _{B})<\ln 2$ and $S(\rho
_{C})<\ln 2$.

Family 7.2. Let Family 7.2 consist of the black holes corresponding to the
following states:

\begin{equation}
|T_{2}\rangle =\eta _{0}|000\rangle -|\eta _{1}||100\rangle +|\eta
_{2}||101\rangle +|\eta _{3}||110\rangle +|\eta _{4}||111\rangle ,
\end{equation}%
where

\begin{eqnarray}
2[(\mu |\eta _{3}|)^{2}+(\mu |\eta _{4}|)^{2}] &=&1, \label{cd-7-2-1} \\
|\eta _{3}| &>&|\eta _{2}|, \label{cd-7-2-2} \\
|\eta _{1}| &=&\frac{|\eta _{2}||\eta _{4}|}{|\eta _{3}|}. \label{cd-7-2-3}
\end{eqnarray}

\ignore{
\begin{eqnarray}
\left( p^{1}p^{3}+p^{2}q_{0}\right)
((p^{1})^{2}-(p^{2})^{2}-q_{0}^{2}+(p^{3})^{2}) &=&0,  \label{cd-7-2-1} \\
((p^{1})^{2}-(p^{2})^{2})((p^{3})-q_{0}^{2}) &<&0,  \label{cd-7-2-2} \\
((p^{1})^{2}-(p^{2})^{2}-q_{0}^{2}+(p^{3})^{2})(p^{1}q_{0}^{3}-(p^{1})^{3}q_{0}+p^{2}(p^{3})^{3}-(p^{2})^{3}p^{3}) &=&0.
\label{cd-7-2-3}
\end{eqnarray}
}
$\mu |T_{2}\rangle $ is normal and

\begin{equation}
\mu |T_{2}\rangle =\mu \eta _{0}|000\rangle -\mu |\eta _{1}||100\rangle +\mu
|\eta _{2}||101\rangle +\mu |\eta _{3}||110\rangle +\mu |\eta
_{4}||111\rangle ,
\end{equation}
where $2[(\mu |\eta _{3}|)^{2}+(\mu |\eta _{4}|)^{2}]=1$ by the condition in
Eq. (\ref{cd-7-2-1}), $|\mu \eta _{3}|>|\mu \eta _{2}|$\ by the condition in
Eq. (\ref{cd-7-2-2}), and $|\eta _{1}|=\frac{|\eta _{2}||\eta _{4}|}{|\eta
_{3}|}$ by the condition in Eq. (\ref{cd-7-2-3}).

By Corollary 5 in Appendix C, $\mu |T_{2}\rangle $ has the maximal von
Neumann entanglement entropy $S(\rho _{B})(=\ln 2)$ while $S(\rho _{A})<\ln
2 $ and $S(\rho _{C})<\ln 2$. Thus, under the factor $\mu $, each black hole
of Family 7.2 has the maximal von Neumann entanglement entropy $S(\rho
_{B})(=\ln 2)$ while $S(\rho _{A})<\ln 2$ and $S(\rho _{C})<\ln 2$.

$\allowbreak $Family 7.3. Let Family 7.3 consist of the black holes
corresponding to the following states:

\begin{equation}
|T_{3}\rangle =\eta _{0}|000\rangle -|\eta _{1}||100\rangle +|\eta
_{2}||101\rangle +|\eta _{3}||110\rangle +|\eta _{4}||111\rangle ,
\end{equation}%
where
\begin{eqnarray}
2[(\mu |\eta _{2}|)^{2}+(\mu |\eta _{4}|)^{2}] &=&1, \label{cd-7-3-1} \\
|\eta _{2}| &>&|\eta _{3}|,  \label{cd-7-3-2}\\
|\eta _{1}| &=&\frac{|\eta _{3}||\eta _{4}|}{|\eta _{2}|}.
\label{cd-7-3-3}
\end{eqnarray}

\ignore{
\begin{eqnarray}
\left( p^{1}q_{0}+p^{2}p^{3}\right)
((p^{1})^{2}-(p^{2})^{2}+q_{0}^{2}-(p^{3})^{2}) &=&0,  \label{cd-7-3-1} \\
((p^{1})^{2}-(p^{2})^{2})((p^{3})-q_{0}^{2}) &>&0,  \label{cd-7-3-2} \\
((p^{1})^{2}-(p^{2})^{2}+q_{0}^{2}-(p^{3})^{2})(p^{1}(p^{3})^{3}+p^{2}q_{0}^{3}-(p^{1})^{3}p^{3}-(p^{2})^{3}q_{0}) &=&0.
\label{cd-7-3-3}
\end{eqnarray}
}
$\mu |T_{3}\rangle $ is normal and

\begin{equation}
\mu |T_{3}\rangle =\mu \eta _{0}|000\rangle -\mu |\eta _{1}||100\rangle +\mu
|\eta _{2}||101\rangle +\mu |\eta _{3}||110\rangle +\mu |\eta
_{4}||111\rangle ,
\end{equation}%
where $2[(\mu |\eta _{2}|)^{2}+(\mu |\eta _{4}|)^{2}]=1$\ by\ the condition
in Eq. (\ref{cd-7-3-1}), $|\mu \eta _{2}|>|\mu \eta _{3}|$\ by\ the
condition in Eq. (\ref{cd-7-3-2}), and $|\mu \eta _{1}|=\frac{|\mu \eta
_{3}||\mu \eta _{4}|}{|\mu \eta _{2}|}$\ \ by the condition in Eq. (\ref{cd-7-3-3}).

By Corollary 6 in Appendix C, $\mu |T_{3}\rangle $ has the maximal von
Neumann entanglement entropy $S(\rho _{C})(=\ln 2)$ while $S(\rho _{A})<\ln 2
$ and $S(\rho _{C})<\ln 2$. Thus, under the factor $\mu $, each black hole
of Family 7.3 has the maximal von Neumann entanglement entropy $S(\rho
_{C})(=\ln 2)$ while $S(\rho _{A})<\ln 2$ and $S(\rho _{B})<\ln 2$.

$\allowbreak $Family 7.4 includes other black holes. Then, by Lemmas 1-8 in
Appendix C each black block of Family 7.4 has von Neumann entanglement
entropy $S(\rho _{x})<\ln 2$, $x=A,B,C$.

{\bf Remark 4.} STU black holes can also be partitioned into eight types by von
Neumann entanglement entropy in Table 4. Clearly, any two types are LU
inequivalent because they have different von Neumann entanglement entropy
which is LU invariant. Therefore, we obtain another LU\ classification of
STU black holes.

\begin{table}[!htbp]
\label{tbl:stubh}
\centering
\caption{Eight types of STU black holes.}
\begin{tabular}{|l|l|l|l|l|}
\toprule
Type  &  & $S(\rho _{A})$ & $S(\rho _{B})$ & $S(\rho _{C})$ \\ \midrule
1 & is Family 1 & $=\ln 2$ & $=\ln 2$ & $=\ln 2$ \\ \midrule
2 & is Family 2 & $=\ln 2$ & $<\ln 2$ & $=\ln 2$ \\ \midrule
3 & is Family 3 & $=\ln 2$ & $=\ln 2$ & $<\ln 2$ \\ \midrule
4 & is Family 4 & $<\ln 2$ & $=\ln 2$ & $=\ln 2$ \\ \midrule
5 & is Family 7.1 & $=\ln 2$ & $<\ln 2$ & $<\ln 2$ \\ \midrule
6 & is Family 7.2 & $<\ln 2$ & $=\ln 2$ & $<\ln 2$ \\ \midrule
7 & is Family 7.3 & $<\ln 2$ & $<\ln 2$ & $=\ln 2$ \\ \midrule
8 & is Families 5, 6, and 7.4 & $<\ln 2$ & $<\ln 2$ & $<\ln 2$ \\ \bottomrule
\end{tabular}
\end{table}

\section{Summary}
\label{sec:summary}
The black holes with $p^{1}p^{2}p^{3}q_{0}\neq 0$ correspond to the
three-qubit states $|\psi \rangle $ in Eq. \eqref{ghz-1}. A complicated
calculation yields the Schmidt decomposition of $\mu |\psi \rangle $ in Eq. \eqref{ghz-1}.
Using the criteria $\frac{(p^{1})^{2}+(p^{2})^{2}}{(p^{3})^{2}+q_{0}^{2}}=1$%
, $\frac{|p^{1}|}{|p^{3}|}=\frac{|p^{2}|}{|q_{0}|}$, and $\frac{|p^{2}|}{%
|p^{3}|}=\frac{|p^{1}|}{|q_{0}|}$ (obtained from $%
(p^{1})^{2}+(p^{2})^{2}-(p^{3})^{2}-q_{0}^{2}=0$, $|p^{2}p^{3}|=|p^{1}q_{0}|$%
, and $|p^{1}p^{3}|=|p^{2}q_{0}|$), the SD of $\mu |\psi \rangle $ in Eq. %
\eqref{ghz-1} is classified into seven groups, and accordingly, the black
holes are classified into seven families. The seven families of black holes
are regarded inequivalent under LU because their corresponding states are
inequivalent under LU.

Clearly, the criteria are independent of the sign of charges. Thus, the
classification of the black holes is not related to the signs of charges,
but related to the ratio of the absolute values of charges. The fact that
the criteria are independent of the signs of charges verifies that between
two black holes with $p^{1}p^{2}p^{3}q_{0}\neq 0$ if their only difference
is the signs of charges, then the two corresponding states are LU equivalent.

In \cite{Linde}, Kallosh and Linde studied the classification of black holes
under SLOCC. Also, in\ \cite{Linde}, Kallosh and Linde claimed that the
theory of stringy black holes requires a more detailed classification than
the standard three-qubit classification provided by their Table 1. Our
contribution is that we do a more detailed classification for the black
holes corresponding to GHZ SLOCC\ class of three qubits and classify the
black holes into seven families under LU. In addition, we also derive the black holes with the maximal von Neumann entanglement entropy and propose another LU classification of black holes by von Neumann entanglement entropy.

As indicated in \cite{Duff}, the appearance of the Cayley hyperdeterminant in the contexts of
stringy black hole entropy and the three-qubit quantum entanglement may be a purely mathematical coincidence, so the intriguing relation between STU extremal black holes and three-qubit systems in quantum information theory may be coincidental. It was pointed out in \cite{Linde} that the coincidence may be a consequence of the underlying symmetry of the theory, i.e., the two different theories have the same underlying symmetry: the symmetry of extremal STU black holes is $[SL(2, \mathbb{R})]^3$; in ABC system the symmetry is $[SL(2, \mathbb{C})]^3$.

 \section*{Data Availability Statement}
 The datasets generated during and/or analyzed during the current study are available from the corresponding author on reasonable request.
\section*{Conflict of interest} The authors have no conflicts of interest to declare that are relevant to the content of this article.

\begin{appendices}

\section{Detailed derivation for the criteria of the seven families}
\label{sec:proof}

{\it Family 1.} Let Family 1 consist of the black holes satisfying the following
Eq. (\ref{ap-cond-1}): 
\begin{equation}
(p^{1})^{2}=(p^{2})^{2}=(p^{3})^{2}=q_{0}^{2}.  \label{ap-cond-1}
\end{equation}

Condition 1. Eq. (\ref{ap-cond-1}) holds if and only if $%
|p^{1}p^{3}|=|p^{2}q_{0}|$, $|p^{2}p^{3}|=|p^{1}q_{0}|$ and $%
(p^{1})^{2}+(p^{2})^{2}-(p^{3})^{2}-q_{0}^{2}=0$.

Proof. 

$(=>)$ If $(p^{1})^{2}=(p^{2})^{2}=(p^{3})^{2}=q_{0}^{2}$, certainly 
$|p^{1}p^{3}|=|p^{2}q_{0}|$, $|p^{2}p^{3}|=|p^{1}q_{0}|$ and $(p^{1})^{2}+(p^{2})^{2}-(p^{3})^{2}-q_{0}^{2}=0$.

$(<=)$ From $|p^{1}p^{3}|=|p^{2}q_{0}|$ and $|p^{2}p^{3}|=|p^{1}q_{0}|$,
obtain $(p^{1})^{2}=(p^{2})^{2}$ and $(p^{3})^{2}=q_{0}^{2}$. Then, from $%
(p^{1})^{2}=(p^{2})^{2}$, $(p^{3})^{2}=q_{0}^{2}$ and $%
(p^{1})^{2}+(p^{2})^{2}-(p^{3})^{2}-q_{0}^{2}=0$, obtain $%
(p^{1})^{2}=q_{0}^{2}$. Thus, obtain Eq. (\ref{ap-cond-1}).

\vspace{5pt}

\noindent {\it Family 2.} Let Family 2 consist of the black holes satisfying the following
Eq. (\ref{ap-cond-2}): 
\begin{equation}
(p^{1})^{2}=(p^{3})^{2},(p^{2})^{2}=q_{0}^{2},(p^{1})^{2}\neq (p^{2})^{2}.
\label{ap-cond-2}
\end{equation}%
\ \ \ Condition 2. Eq. (\ref{ap-cond-2}) holds if and only if $%
(p^{1})^{2}+(p^{2})^{2}-(p^{3})^{2}-q_{0}^{2}=0$, $|p^{2}p^{3}|=|p^{1}q_{0}|$%
, but $|p^{1}p^{3}|\neq |p^{2}q_{0}|$.

Proof. 

$(=>)$ Clearly, $(p^{1})^{2}+(p^{2})^{2}-(p^{3})^{2}-q_{0}^{2}=0$
and $|p^{2}p^{3}|=|p^{1}q_{0}|$ . Assume that $|p^{1}p^{3}|=|p^{2}q_{0}|$ .
In light of Condition 1, we get $%
(p^{1})^{2}=(p^{2})^{2}=(p^{3})^{2}=q_{0}^{2}$, which contradicts $%
(p^{1})^{2}\neq (p^{2})^{2}$.

$(<=)$ From $|p^{2}p^{3}|=|p^{1}q_{0}|$, let $\frac{\left\vert
p^{2}\right\vert }{\left\vert q_{0}\right\vert }=\frac{\left\vert
p^{1}\right\vert }{\left\vert p^{3}\right\vert }=\ell $. Then, $\left\vert
p^{2}\right\vert =\ell \left\vert q_{0}\right\vert $ and $\left\vert
p^{1}\right\vert =\ell \left\vert p^{3}\right\vert $. Then, $%
(p^{1})^{2}+(p^{2})^{2}-(p^{3})^{2}-q_{0}^{2}=(\ell
^{2}-1)[(p^{3})^{2}+q_{0}^{2}]$. Let $(\ell ^{2}-1)[(p^{3})^{2}+q_{0}^{2}]=0$%
. Then, we get $\ell =1$. Thus, we get $(p^{1})^{2}=(p^{3})^{2}$ and $%
(p^{2})^{2}=q_{0}^{2}$. Then, from $|p^{1}p^{3}|\neq |p^{2}q_{0}|$ we obtain 
$(p^{1})^{2}\neq (p^{2})^{2}$.

\vspace{5pt}

\noindent {\it Family 3.} Let Family 3 consist of the black holes satisfying the following
Eq. (\ref{ap-cond-3}):

\begin{equation}
(p^{2})^{2}=(p^{3})^{2},(p^{1})^{2}=q_{0}^{2},(p^{1})^{2}\neq (p^{2})^{2}
\label{ap-cond-3}
\end{equation}

\ \ Condition 3. Eq. (\ref{ap-cond-3}) holds if and only if $%
(p^{1})^{2}+(p^{2})^{2}-(p^{3})^{2}-q_{0}^{2}=0$, $|p^{1}p^{3}|=|p^{2}q_{0}|$%
, but $|p^{2}p^{3}|\neq |p^{1}q_{0}|$.

Proof. 

$(=>)$ Clearly, $(p^{1})^{2}+(p^{2})^{2}-(p^{3})^{2}-q_{0}^{2}=0$
and $|p^{1}p^{3}|=|p^{2}q_{0}|$. Assume that $|p^{2}p^{3}|=|p^{1}q_{0}|$. In
light of Condition 1, we obtain $%
(p^{1})^{2}=(p^{2})^{2}=(p^{3})^{2}=q_{0}^{2}$, which contradicts $%
(p^{1})^{2}\neq (p^{2})^{2}$.

$(<=)$ Via $|p^{1}p^{3}|=|p^{2}q_{0}|$, let $\frac{\left\vert
p^{2}\right\vert }{\left\vert p^{3}\right\vert }=\frac{\left\vert
p^{1}\right\vert }{\left\vert q_{0}\right\vert }=\ell $. Then, $\left\vert
p^{2}\right\vert =\ell \left\vert p^{3}\right\vert $ and $\left\vert
p^{1}\right\vert =\ell \left\vert q_{0}\right\vert $. Then, $%
(p^{1})^{2}+(p^{2})^{2}-(p^{3})^{2}-q_{0}^{2}=(\ell
^{2}-1)[(p^{3})^{2}+q_{0}^{2}]$. Let $(\ell ^{2}-1)[(p^{3})^{2}+q_{0}^{2}]=0$%
. Then, we get $\ell =1$. Thus, we get $(p^{1})^{2}=q_{0}^{2}$, and $%
(p^{2})^{2}=(p^{3})^{2}$. Then, from $|p^{2}p^{3}|\neq |p^{1}q_{0}|$, we
obtain $(p^{1})^{2}\neq (p^{2})^{2}$.

\vspace{5pt}

\noindent {\it Family 4.} Let Family 4 consist of the black holes satisfying the following
Eq. (\ref{ap-cond-4}): 
\begin{equation}
(p^{1})^{2}=(p^{2})^{2},(p^{3})^{2}=q_{0}^{2},(p^{1})^{2}\neq (p^{3})^{2}
\label{ap-cond-4}
\end{equation}%
Condition 4. Eq. (\ref{ap-cond-4}) holds if and only if $%
(p^{1})^{2}+(p^{2})^{2}-(p^{3})^{2}-q_{0}^{2}\neq 0$, but $%
|p^{2}p^{3}|=|p^{1}q_{0}|$, and $|p^{1}p^{3}|=|p^{2}q_{0}|$.

Proof. 

$(=>)$ It is trivial.

$(<=)$ From $|p^{2}p^{3}|=|p^{1}q_{0}|$ and $|p^{1}p^{3}|=|p^{2}q_{0}|$, we
get $(p^{1})^{2}=(p^{2})^{2}$and $(p^{3})^{2}=q_{0}^{2}$. Then, from $%
(p^{1})^{2}+(p^{2})^{2}-(p^{3})^{2}-q_{0}^{2}=2[(p^{1})^{2}-(p^{3})^{2}]\neq
0$, we get $(p^{1})^{2}\neq (p^{3})^{2}$.

\vspace{5pt}

\noindent {\it Family 5.} Let Family 5 consist of the black holes satisfying the following
Eq. (\ref{ap-cond-5}): 
\begin{equation}
(p^{2})^{2}\neq (p^{3})^{2},(p^{3})^{2}\neq
q_{0}^{2},|p^{1}p^{3}|=|p^{2}q_{0}|  \label{ap-cond-5}
\end{equation}%
Condition 5. Eq. (\ref{ap-cond-5}) holds if and only if $%
(p^{1})^{2}+(p^{2})^{2}-(p^{3})^{2}-q_{0}^{2}\neq 0,|p^{2}p^{3}|\neq
|p^{1}q_{0}|,|p^{1}p^{3}|=|p^{2}q_{0}|$.

Proof. From $|p^{1}p^{3}|=|p^{2}q_{0}|$, let $\frac{|p^{1}|}{|q_{0}|}=\frac{%
|p^{2}|}{|p^{3}|}=\ell $. Then, 
\begin{eqnarray}
|p^{2}| &=&\ell |p^{3}|,  \label{result-5-1} \\
|p^{1}| &=&\ell |q_{0}|.  \label{result-5-2}
\end{eqnarray}

$(<=)$ From Eqs. (\ref{result-5-1}, \ref{result-5-2}) and $%
(p^{1})^{2}+(p^{2})^{2}-(p^{3})^{2}-q_{0}^{2}\neq 0$, we obtain $\ell \neq 1$%
. Thus, $|p^{2}|\neq |p^{3}|$ and $|p^{1}|\neq |q_{0}|$. Via $%
|p^{2}p^{3}|\neq |p^{1}q_{0}|$ and Eqs. (\ref{result-5-1}, \ref{result-5-2}%
), we obtain $(p^{3})^{2}\neq q_{0}^{2}$.

$(=>)$ Via $(p^{2})^{2}\neq (p^{3})^{2}$ and Eq. (\ref{result-5-1}), obtain 
$\ell \neq 1$. Via Eq. (\ref{result-5-1}) and $(p^{3})^{2}\neq q_{0}^{2}$,
obtain $|p^{2}||p^{3}|=\ell |p^{3}|^{2}$ $\neq \ell q_{0}^{2}$. Via Eq. ( %
\ref{result-5-2}), obtain $|p^{1}||q_{0}|=\ell q_{0}^{2}$. Thus, obtain $%
|p^{2}p^{3}|\neq |p^{1}q_{0}|$. Via $\ell \neq 1$ and Eqs. (\ref{result-5-1}%
, \ref{result-5-2}), obtain $(p^{1})^{2}+(p^{2})^{2}-(p^{3})^{2}-q_{0}^{2}%
\neq 0$.

\vspace{5pt}

\noindent {\it Family 6.} Let Family 6 consist of the black holes satisfying the following
Eq. (\ref{ap-cond-6}): 
\begin{equation}
(p^{2})^{2}\neq q_{0}^{2},(p^{3})^{2}\neq
q_{0}^{2},|p^{2}p^{3}|=|p^{1}q_{0}|.  \label{ap-cond-6}
\end{equation}%
Condition 6. Eq. (\ref{ap-cond-6}) holds if and only if $%
(p^{1})^{2}+(p^{2})^{2}-(p^{3})^{2}-q_{0}^{2}\neq
0,|p^{2}p^{3}|=|p^{1}q_{0}|,|p^{1}p^{3}|\neq |p^{2}q_{0}|$.

Proof. From $|p^{2}p^{3}|=|p^{1}q_{0}|$, let $\frac{|p^{2}|}{|q_{0}|}=\frac{%
|p^{1}|}{|p^{3}|}=\ell $. Then, 
\begin{eqnarray}
|p^{2}| &=&\ell |q_{0}|,  \label{result-6-1} \\
|p^{1}| &=&\ell |p^{3}|.  \label{result-6-2}
\end{eqnarray}

$(<=)$ From Eqs. (\ref{result-6-1}, \ref{result-6-2}) and $%
(p^{1})^{2}+(p^{2})^{2}-(p^{3})^{2}-q_{0}^{2}\neq 0$, obtain $\ell \neq 1$.
Then, obtain $|p^{2}|\neq |q_{0}|$ and $|p^{1}|\neq |p^{3}|$. Then, from
Eqs. (\ref{result-6-1}, \ref{result-6-2}) and $|p^{1}p^{3}|\neq |p^{2}q_{0}|$%
, obtain $(p^{3})^{2}\neq q_{0}^{2}$.

$(=>)$ Via Eq. (\ref{result-6-1}) and $(p^{2})^{2}\neq q_{0}^{2}$, obtain $%
\ell \neq 1$. Via $\ell \neq 1$ and Eqs. (\ref{result-6-1}, \ref{result-6-2}%
), obtain $(p^{1})^{2}+(p^{2})^{2}-(p^{3})^{2}-q_{0}^{2}\neq 0$. Via Eqs. (%
\ref{result-6-1}, \ref{result-6-2}), $|p^{1}p^{3}|=\ell (p^{3})^{2}$ and $%
|p^{2}q_{0}|=\ell q_{0}^{2}$. Since $(p^{3})^{2}\neq q_{0}^{2}$, then $\ell
(p^{3})^{2}\neq \ell q_{0}^{2}$, and $|p^{1}p^{3}|\neq |p^{2}q_{0}|$.

{
\vspace{5pt}

\noindent {\it Family 7.} Let Family 7 consist of the black holes satisfying the following
Eq. (\ref{ap-cond-7}): 

\begin{equation}
\frac{|p^{1}|}{|q_{0}|}\neq \frac{|p^{2}|}{|p^{3}|}, \frac{|p^{2}|}{|q_{0}|}\neq \frac{|p^{1}|}{|p^{3}|}. \label{ap-cond-7}
\end{equation}

Condition 7. Eq. (\ref{ap-cond-7}) holds if none of the other six conditions holds, so that the seven families form a partition of the space based on the equivalence classes of black hole charges.
}

The seven families and their criteria are summarized in Table \ref{tbl:family} in the main text.

\section{The 16 Dictionaries}
\label{dictionary}

\subsection{Duff's correspondence between charges and local bases product states has 16 dictionaries}

The entropy of black holes can be shortened. We know that to make a relation between the
entropy of black holes and 3-tangle $\tau _{ABC}$, it is necessary to make $\Delta =\det $ $\Psi $, where $\Delta$ is defined as in eq. \eqref{entropy-1}.

To make $\Delta =\det $ $\Psi $, one way
is to associate all magnetic charges (electric charges) with the presence of 1's (0's) in local bases product states from the rightmost position to the leftmost position. That is, let the state with magnetic charge $p^{0}$ ($p^{1}$, $p^{2}$, $
p^{3}$) correspond to $|000\rangle $ ($|001\rangle $, $|010\rangle $, $
|100\rangle $), and the state with electric charge $q_{0}$ ($q_{1}$, $q_{2}
$, $q_{3}$) correspond to $|111\rangle $ ($|110\rangle $, $|101\rangle $, $
|011\rangle $) \cite{Duff} (see Table \ref{tbl:app-tbl1}).

\begin{table}[!htbp]
\caption{Duff's correspondence between charges and local bases product states.}
\label{tbl:app-tbl1}
\begin{tabular}{|l|c|c|c|c|c|c|c|c|}
\toprule
Charge& $p^{0}$ & $q_{3}$ & $p^{1}$ & $p^{2}$ & $q_{0}$ & $p^{3}$ & $q_{1}$ & $q_{2}$\\ 
\midrule
State &$|000\rangle $ & $|011\rangle $ & $|001\rangle $ & $|010\rangle $ & $|111\rangle $ & $|100\rangle $ & $|110\rangle $ & $|101\rangle $ \\ 
\bottomrule
\end{tabular}
\end{table}

Table \ref{tbl:app-tbl1} gives the correspondence between the charges and the states, however, the signs of the charges are not explained. To determine the signs, let $\delta _{i} \in \{ +1, -1\}$, and we then create Table \ref{tbl:app-tbl2}. By substituting the charges in $\Delta $ with the corresponding
items in Table \ref{tbl:app-tbl2} and solving the equation $\Delta =\det $ $\Psi $, we obtain all the 16 solutions for $\delta _{i}$. Thus, we obtain 16 different versions of Table
\ref{tbl:app-tbl2}. Each version of the table is called a dictionary \cite{Linde}. In total, there are 16 dictionaries for Duff's correspondence. The 16 dictionaries are generated by having $B=\pm C_{i}$ for $i=1, \ldots, 8$.  Table \ref{tbl:app-tbl3} shows eight dictionaries generated by having $B=C_i$, and another eight dictionaries can be generated by having $B=-C_i$, in which all elements in the table will have their signs flipped. In the literature, the dictionaries generated by having $B=C_{i}$ with $i=1,2,3$ appeared in \cite{Duff, Linde, Levay}.

\begin{table}[!htbp]
\caption{The general dictionary.}
\label{tbl:app-tbl2}
\begin{tabular}{|m{2cm} | m{2cm}|}
\toprule
$p^{0}$ & $\delta _{0}a_{000}$ \\ \midrule
$p^{1}$ & $\delta _{1}a_{001}$ \\ \midrule
$p^{2}$ & $\delta _{2}a_{010}$ \\ \midrule
$p^{3}$ & $\delta _{4}a_{100}$ \\ \midrule
$q_{0}$ & $\delta _{7}a_{111}$ \\ \midrule
$q_{1}$ & $\delta _{6}a_{110}$ \\ \midrule
$q_{2}$ & $\delta _{5}a_{101}$ \\ \midrule
$q_{3}$ & $\delta _{3}a_{011}$ \\ \bottomrule
\end{tabular}
\end{table}

\begin{table}[!htbp]
\caption{Eight dictionaries generated by having $B=C_i$, $i=1, \ldots, 8$.}
\label{tbl:app-tbl3}
\begin{tabular}{|l|l|l|l|l|l|l|l|l|}
\toprule
$B$ & $C_{1}$ & $C_{2}$ & $C_{3}$ & $C_{4}$ & $C_{5}$ & $C_{6}$ & $C_{7}$ & $
C_{8}$ \\ \midrule
$p^{0}$ & $-a_{000}$ & $a_{000}$ & $a_{000}$ & $a_{000}$ & $a_{000}$ & $
-a_{000}$ & $-a_{000}$ & $-a_{000}$ \\ \midrule
$p^{1}$ & $-a_{001}$ & $-a_{001}$ & $a_{001}$ & $-a_{001}$ & $a_{001}$ & $
a_{001}$ & $-a_{001}$ & $a_{001}$ \\ \midrule
$p^{2}$ & $-a_{010}$ & $-a_{010}$ & $a_{010}$ & $a_{010}$ & $-a_{010}$ & $
a_{010}$ & $a_{010}$ & $-a_{010}$ \\ \midrule
$p^{3}$ & $a_{100}$ & $-a_{100}$ & $a_{100}$ & $a_{100}$ & $a_{100}$ & $
-a_{100}$ & $a_{100}$ & $a_{100}$ \\ \midrule
$q_{0}$ & $-a_{111}$ & $a_{111}$ & $-a_{111}$ & $a_{111}$ & $a_{111}$ & $
a_{111}$ & $a_{111}$ & $a_{111}$ \\ \midrule
$q_{1}$ & $a_{110}$ & $a_{110}$ & $a_{110}$ & $a_{110}$ & $-a_{110}$ & $
a_{110}$ & $-a_{110}$ & $a_{110}$ \\ \hline
$q_{2}$ & $a_{101}$ & $a_{101}$ & $a_{101}$ & $-a_{101}$ & $a_{101}$ & $
a_{101}$ & $a_{101}$ & $-a_{101}$ \\ \midrule
$q_{3}$ & $-a_{011}$ & $a_{011}$ & $a_{011}$ & $-a_{011}$ & $-a_{011}$ & $
-a_{011}$ & $a_{011}$ & $a_{011}$ \\ \bottomrule
\end{tabular}
\end{table}

\subsection{There are many correspondences between the charges and local
bases product states different from Duff's. }
A correspondence must satisfy $\Delta =\det $ $\Psi $. It is easy to see that to ensure $\Delta =\det $ $\Psi $, when the state with magnetic charge $p^{i}$, $i=0,1,2,3$, correspond to the state $|i_{3}i_{2}i_{1}\rangle $, where $i_1, i_2, i_3 \in \{0, 1\}$, then $q_{i}$\ must correspond to $|\tilde{\imath}_{3}\tilde{\imath}_{2}\tilde{\imath}_{1}\rangle $, where $\tilde{\imath}_{\ell }=1-i_{\ell }$, $\ell =1,2,3$, via eqs. \eqref{entropy-1} and \eqref{Cayley}. From this property we can construct many correspondences. For example,  by exchanging $p^{1}$ and $p^{2}$ in Duff's correspondence, we can obtain another correspondence.

\section{The states with the maximal von Neumann entanglement entropy}
\label{maxvne}

It is well known that von Neumann entanglement entropy is an important
entanglement measure. We explore what states have the maximal von Neumann
entanglement entropy. For readability, we list the above Schmidt
decomposition for the non-BPS black holes with $p^{1}p^{2}p^{3}q_{0}<0$ as
follows. 
\begin{equation}
\lambda _{0}|000\rangle \pm \lambda _{1}|100\rangle +\lambda _{2}|101\rangle
+\lambda _{3}|110\rangle +\lambda _{4}|111\rangle ,
\end{equation}%
where $\lambda _{i}\geq 0$, $\sum_{i=0}^{4}\lambda _{i}^{2}=1$, the state is
written as $(\lambda _{0},\pm \lambda _{1},\lambda _{2},\lambda _{3},\lambda
_{4})$ sometimes. It is well known that $\lambda _{0}\lambda _{4}\neq 0$ for
GHZ SLOCC class. We use the notations $J_{1}=|\pm \lambda _{1}\lambda
_{4}-\lambda _{2}\lambda _{3}|^{2}$, $J_{i}=(\lambda _{0}\lambda _{i})^{2}$, 
$i=2,3,4$, where $J_{i}$, $i=1,2,3,4$, are LU invariant and defined in \cite%
{Acin00}. 

It is known that\ $\rho _{A}=tr_{BC}\rho _{ABC}$, where $\rho _{ABC}$ is the
density matrix. For the states of the form $(\lambda _{0},\pm \lambda
_{1},\lambda _{2},\lambda _{3},\lambda _{4})$, von Neumann entanglement
entropy $S(\rho _{x})$,\ where $x=A,B,C$, is given as follows \cite{dli-qip}%
.  
\begin{equation}
\resizebox{.9 \textwidth}{!} { $ S(\rho _{x })=-\left(
\frac{1+\sqrt{1-4\alpha _{x }}}{2}\ln \frac{1+\sqrt{1-4\alpha _{x
}}}{2}+\frac{1-\sqrt{1-4\alpha _{x }}}{2}\ln \frac{1-\sqrt{1-4\alpha _{x
}}}{2}\right) $ }
\end{equation}%
where $0\leq \alpha _{x}\leq 1/4$. (see Table \ref{tbl:alpha}.) We showed
that $S(\rho _{x})$ increases strictly monotonically as $\alpha _{x}$
increases. Thus, $S(\rho _{x})=\ln 2$ if and only if $\alpha _{x}=1/4$. It
is well known that the maximal von Neumann entanglement entropy is $\ln 2$.

\begin{table}[!tbph]
\caption{Values of $\protect\alpha_{A}$,$\protect\alpha_{B}$, and $\protect\alpha _{C}$.}
\label{tbl:alpha}\centering
\begin{tabular}{|l|l|}
\toprule $\alpha _{A}$ & $J_{2}+J_{3}+J_{4}$ \\ 
\midrule $\alpha _{B}$ & $J_{1}+J_{3}+J_{4}$ \\ 
\midrule $\alpha _{C}$ & $J_{1}+J_{2}+J_{4}$ \\ 
\bottomrule  
\end{tabular}
\end{table}

{\bf Lemma 1.} A state of the form $(\lambda _{0},\pm \lambda _{1},\lambda
_{2},\lambda _{3},\lambda _{4})$ has the maximal von Neumann entanglement
entropy $S(\rho _{A})=S(\rho _{B})=S(\rho _{C})(=\ln 2)$ if and only if the
state is GHZ state $\frac{1}{\sqrt{2}}(|000\rangle +|111\rangle )$.

From \cite{dli-qip}, it is easy to see that Lemma 1 holds.

{\bf Lemma 2.} $\alpha _{A}=1/4$ implies $\lambda _{1}=0$ and $\lambda _{0}=1/%
\sqrt{2}$.

Proof. By the definition of $\alpha _{A}$, we have the following equation. 
\begin{equation}
\alpha _{A}=J_{2}+J_{3}+J_{4}=1/4.  \label{entropy-a-1}
\end{equation}

From Eq. (\ref{entropy-a-1}), obtain 
\begin{equation}
\lambda _{0}^{2}(\lambda _{2}^{2}+\lambda _{3}^{2}+\lambda _{4}^{2})=\lambda
_{0}^{2}(1-\lambda _{0}^{2}-\lambda _{1}^{2})=1/4,  \label{ent-a-4}
\end{equation}%
and then 
\begin{equation}
\lambda _{0}^{4}-\lambda _{0}^{2}(1-\lambda _{1}^{2})+1/4=0.
\end{equation}

The above equation has solutions for $\lambda _{0}^{2}$ if and only if the
discriminant $\ (1-\lambda _{1}^{2})^{2}-1\geq 0$, i.e., $(1-\lambda
_{1}^{2})^{2}\geq 1$. Clearly, $(1-\lambda _{1}^{2})^{2}\geq 1$ if and only
if $\lambda _{1}=0$. Then, obtain 
\begin{equation}
\lambda _{1}=0,\lambda _{0}=1/\sqrt{2}  \label{ent-a-5}
\end{equation}

{\bf Lemma 3.} The state of the form $(\lambda _{0},\pm \lambda _{1},\lambda
_{2},\lambda _{3},\lambda _{4})$ has the maximal von Neumann entanglement
entropy $S(\rho _{A})=S(\rho _{C})=\ln 2$ if and only if the state is $\frac{%
1}{\sqrt{2}}|000\rangle +\lambda _{2}|101\rangle +\lambda _{4}|111\rangle ,$

Proof. Since $S(\rho _{A})=S(\rho _{C})=\ln 2$, then we have the following
equations:

\begin{eqnarray}
\alpha _{A} &=&J_{2}+J_{3}+J_{4}=1/4,  \label{ent-c-0} \\
\alpha _{C} &=&J_{1}+J_{2}+J_{4}=1/4.
\end{eqnarray}%
Then, from the above two equations, obtain 
\begin{equation}
J_{1}=J_{3}.  \label{ent-c-1}
\end{equation}

From Eq. (\ref{ent-c-0}) and by Lemma 2, obtain 
\begin{equation}
\lambda _{1}=0,\lambda _{0}=1/\sqrt{2}  \label{ent-c-2}
\end{equation}

From Eqs. (\ref{ent-c-1}, \ref{ent-c-2}), 
\begin{equation}
\lambda _{2}\lambda _{3}=\lambda _{0}\lambda _{3}=\frac{1}{\sqrt{2}}\lambda
_{3}.
\end{equation}

From the above equation, one can see that if $\lambda _{3}\neq 0$ then $%
\lambda _{2}=\frac{1}{\sqrt{2}}$. It is impossible because $%
\sum_{i=0}^{4}\lambda _{i}^{2}=1$. So, $\lambda _{3}=0$. Then, obtain the
following state 
\begin{equation}
|\xi \rangle =\frac{1}{\sqrt{2}}|000\rangle +\lambda _{2}|101\rangle
+\lambda _{4}|111\rangle .
\end{equation}

Thus, we show that if a state of the form $(\lambda _{0},\pm \lambda
_{1},\lambda _{2},\lambda _{3},\lambda _{4})$\ has the maximal von Neumann
entanglement entropy $S(\rho _{A})=S(\rho _{C})$ ($=\ln 2$), then the state
must be $|\xi \rangle $. Conversely, let us calculate $S(\rho _{x})$, $%
x=A,B,C$,\ for the state $|\xi \rangle $ below. Clearly, $J_{1}=0$, $J_{3}=0$%
, $J_{2}=(1/2)\lambda _{3}^{2}$ and $J_{4}=(1/2)\lambda _{4}^{2}$. So, $%
\alpha _{A}=\alpha _{C}=J_{2}+J_{4}=(1/2)(\lambda _{2}^{2}+\lambda
_{4}^{2})=1/4$ and $S(\rho _{A})=S(\rho _{C})=\ln 2$. $\alpha _{B}=\frac{1}{2%
}\lambda _{4}{}^{2}$. Thus, $|\xi \rangle $ has the maximal von Neumann
entanglement entropy $S(\rho _{A})=S(\rho _{C})=\ln 2$.

From Lemmas 1 and 3, we have the following corollary.

{\bf Corollary 1.} The state of the form of $(\lambda _{0},\pm \lambda
_{1},\lambda _{2},\lambda _{3},\lambda _{4})$ has the maximal von Neumann
entanglement entropy $S(\rho _{A})=S(\rho _{C})$ ($=\ln 2$) while $S(\rho
_{B})<\ln 2$ if and only if the state is $\frac{1}{\sqrt{2}}|000\rangle
+\lambda _{2}|101\rangle +\lambda _{4}|111\rangle ,$ where $\lambda _{2}\neq
0$.

{\bf Lemma 4.} The state of the form $(\lambda _{0},\pm \lambda _{1},\lambda
_{2},\lambda _{3},\lambda _{4})$ has the maximal von Neumann entanglement
entropy $S(\rho _{A})=S(\rho _{B})$ ($=\ln 2$) if and only if the state is $%
\frac{1}{\sqrt{2}}|000\rangle +\lambda _{3}|110\rangle +\lambda
_{4}|111\rangle $.

Proof. Since $S(\rho _{A})=S(\rho _{B})=\ln 2$, then we have the following
equations 
\begin{eqnarray}
\alpha _{A} &=&J_{2}+J_{3}+J_{4}=1/4,  \label{ent-b-1} \\
\alpha _{B} &=&J_{1}+J_{3}+J_{4}=1/4.
\end{eqnarray}

Then, from the above obtain 
\begin{equation}
J_{1}=J_{2}.  \label{ent-b-2}
\end{equation}

From Eq. (\ref{ent-b-1}) and by Lemma 2, obtain 
\begin{equation}
\lambda _{1}=0,\lambda _{0}=1/\sqrt{2}.  \label{ent-b-5}
\end{equation}

From Eqs. (\ref{ent-b-2}, \ref{ent-b-5}), 
\begin{equation}
\lambda _{2}\lambda _{3}=\lambda _{0}\lambda _{2}=\frac{1}{\sqrt{2}}\lambda
_{2}.  \label{ent-b-6}
\end{equation}

From Eq. (\ref{ent-b-6}), If $\lambda _{2}\neq 0$, then $\lambda _{3}=\frac{1%
}{\sqrt{2}}$. However, it is impossible because $\sum_{i=0}^{4}\lambda
_{i}^{2}=1$. Therefore, $\lambda _{2}=0$ and then, we obtain 
\begin{equation}
|\varpi \rangle =\frac{1}{\sqrt{2}}|000\rangle +\lambda _{3}|110\rangle
+\lambda _{4}|111\rangle .
\end{equation}

Thus, we show that if a state of the form $(\lambda _{0},\pm \lambda
_{1},\lambda _{2},\lambda _{3},\lambda _{4})$\ has the maximal von Neumann
entanglement entropy $S(\rho _{A})=S(\rho _{B})=\ln 2$, then the state must
be $|\varpi \rangle $. Conversely, let us calculate $S(\rho _{x})$, $x=A,B,C$%
,\ for the state $|\varpi \rangle $ below. Clearly, $J_{1}=0$, $J_{2}=0$, $%
J_{3}=(1/2)\lambda _{3}^{2}$ and $J_{4}=(1/2)\lambda _{4}^{2}$. So, $\alpha
_{A}=\alpha _{B}=J_{3}+J_{4}=(1/2)(\lambda _{3}^{2}+\lambda _{4}^{2})=1/4$
and $S(\rho _{A})=S(\rho _{B})=\ln 2$. $\alpha _{C}=\frac{1}{2}\lambda
_{4}{}^{2}$. Thus, $|\varpi \rangle $ has the maximal von Neumann
entanglement entropy $S(\rho _{A})=S(\rho _{B})=\ln 2$.

From Lemmas 1 and 4, we have the following corollary.

{\bf Corollary 2.} The state of the form $(\lambda _{0},\pm \lambda _{1},\lambda
_{2},\lambda _{3},\lambda _{4})$ has the maximal von Neumann entanglement
entropy $S(\rho _{A})=S(\rho _{B})$ ($=\ln 2$) while $S(\rho _{C})<\ln 2$ if
and only if the state is $\frac{1}{\sqrt{2}}|000\rangle +\lambda
_{3}|110\rangle +\lambda _{4}|111\rangle $, where $\lambda _{3}\neq 0$.

{\bf Lemma 5.} The state of the form $(\lambda _{0},\pm \lambda _{1},\lambda
_{2},\lambda _{3},\lambda _{4})$ has the maximal von Neumann entanglement
entropy $S(\rho _{A})$ ($=ln2$) if and only if the state is $\frac{1}{\sqrt{2%
}}|000\rangle +\lambda _{2}|101\rangle +\lambda _{3}|110\rangle +\lambda
_{4}|111\rangle $.

Proof. That $S(\rho _{A})$ $=\ln 2$ implies $\alpha _{A}=1/4$. By Lemma 2,
obtain 
\begin{equation}
\lambda _{1}=0,\lambda _{0}=1/\sqrt{2}
\end{equation}%
and then%
\begin{equation}
|\kappa \rangle =\frac{1}{\sqrt{2}}|000\rangle +\lambda _{2}|101\rangle
+\lambda _{3}|110\rangle +\lambda _{4}|111\rangle .
\end{equation}

Thus, we show that if a state of the form $(\lambda _{0},\pm \lambda
_{1},\lambda _{2},\lambda _{3},\lambda _{4})$\ has the maximal von Neumann
entanglement entropy $S(\rho _{A})=\ln 2$, then the state must be $|\kappa
\rangle $. Conversely, one can verify that for the state $|\kappa \rangle $, 
$\alpha _{A}=1/4$ and so $S(\rho _{A})=\ln 2$.

Lemmas 1, 3, 4, and 5 imply the following corollary.

{\bf Corollary 3.} The state of the form $(\lambda _{0},\pm \lambda _{1},\lambda
_{2},\lambda _{3},\lambda _{4})$ has the maximal von Neumann entanglement
entropy $S(\rho _{A})$ ($=ln2$) while $S(\rho _{B})<\ln 2$ and $S(\rho
_{C})<\ln 2$ if and only if the state is $\frac{1}{\sqrt{2}}|000\rangle
+\lambda _{2}|101\rangle +\lambda _{3}|110\rangle +\lambda _{4}|111\rangle $%
, where $\lambda _{2}\lambda _{3}\neq 0$.

{\bf Lemma 6.} The state of the form $(\lambda _{0},\pm \lambda _{1},\lambda
_{2},\lambda _{3},\lambda _{4})$ has the maximal von Neumann entanglement
entropy $S(\rho _{B})=S(\rho _{C})$ ($=\ln 2$) if and only if the state is $%
\lambda _{0}|000\rangle \pm \lambda _{1}|100\rangle +\frac{1}{\sqrt{2}}%
|111\rangle $.

Proof. That $S(\rho _{B})=S(\rho _{C})=\ln 2$ implies 
\begin{eqnarray}
\alpha _{B} &=&J_{1}+J_{3}+J_{4}=1/4  \label{a-b} \\
\alpha _{C} &=&J_{1}+J_{2}+J_{4}=1/4  \label{a-c}
\end{eqnarray}

Then from Eqs. (\ref{a-b}, \ref{a-c}), obtain $J_{2}=J_{3}$, and then $%
\lambda _{2}=\lambda _{3}$. From the definition of $\alpha _{B}$, obtain 
\begin{equation}
\lambda _{1}^{2}\lambda _{4}^{2}\pm 2\lambda _{1}\lambda _{2}^{2}\lambda
_{4}+\lambda _{2}^{4}+\lambda _{0}^{2}\lambda _{2}^{2}+\lambda
_{0}^{2}\lambda _{4}^{2}-1/4=0  \label{disc-0}
\end{equation}

Note that $\sum_{i=0}^{4}\lambda _{i}^{2}=1$. $\allowbreak $By substituting $%
\lambda _{0}$ with $(1-(\lambda _{1}^{2}+2\lambda _{2}^{2}+\lambda
_{4}^{2})) $, obtain from Eq. (\ref{disc-0}) 
\begin{equation}
\lambda _{1}^{2}\lambda _{2}^{2}\pm 2\lambda _{1}\lambda _{2}^{2}\lambda
_{4}+\lambda _{2}^{4}+3\lambda _{2}^{2}\lambda _{4}^{2}-\allowbreak \lambda
_{2}^{2}+\lambda _{4}^{4}-\lambda _{4}^{2}+\frac{1}{4}=0  \label{disc-1}
\end{equation}

Case 1. $\lambda _{2}=0$. Eq. (\ref{disc-1}) becomes $\lambda
_{4}^{4}-\lambda _{4}^{2}+\frac{1}{4}=0$.

Then, obtain 
\begin{equation}
\lambda _{4}=\frac{1}{\sqrt{2}}
\end{equation}%
and then 
\begin{equation}
|\varrho \rangle =\lambda _{0}|000\rangle \pm \lambda _{1}|100\rangle +\frac{%
1}{\sqrt{2}}|111\rangle .  \label{state-bc-1}
\end{equation}

Case 2. $\lambda _{2}\neq 0$. We next show that Eq. (\ref{disc-1}) does not
have a solution for $\lambda _{1}$ whenever $\lambda _{2}\neq 0$.

It is clear that the discriminant for $\lambda _{1}$ in Eq. (\ref{disc-1})
is 
\begin{eqnarray}
\Delta &=&(\pm 2\lambda _{2}^{2}\lambda _{4})^{2}-4\lambda _{2}^{2}(\lambda
_{2}^{4}+3\lambda _{2}^{2}\lambda _{4}^{2}-\allowbreak \lambda
_{2}^{2}+\lambda _{4}^{4}-\lambda _{4}^{2}+\frac{1}{4}) \\
&=&-\lambda _{2}^{2}\left( 2\lambda _{2}^{2}+2\lambda _{4}^{2}-1\right) ^{2}.
\label{delta-bc-3}
\end{eqnarray}

Let $\Delta =0$ in Eq. (\ref{delta-bc-3}). Then, obtain $2\lambda
_{2}^{2}+2\lambda _{4}^{2}=1$. Thus, from Eq. (\ref{disc-1}), obtain 
\begin{equation}
\lambda _{1}=\pm \lambda _{4}.
\end{equation}

It is impossible for $\lambda _{1}=-\lambda _{4}$. When $\lambda
_{1}=\lambda _{4}$, from $\sum_{i=0}^{4}\lambda _{i}^{2}=1$ and $2\lambda
_{2}^{2}+2\lambda _{4}^{2}=1$, obtain 
\begin{equation}
\lambda _{0}^{2}=1-(\lambda _{1}^{2}+2\lambda _{2}^{2}+\lambda
_{4}^{2})=1-(2\lambda _{2}^{2}+2\lambda _{4}^{2})=0.
\end{equation}

It is also impossible for $\lambda _{0}=0$ because $\lambda _{0}\lambda
_{4}\neq 0$ for GHZ SLOCC class.

Thus, we show that if a state of the form $(\lambda _{0},\pm \lambda
_{1},\lambda _{2},\lambda _{3},\lambda _{4})$\ has the maximal von Neumann
entanglement entropy $S(\rho _{B})=S(\rho _{C})(=\ln 2)$, then the state
must be $|\varrho \rangle $. Conversely, one can verify that for the state $%
|\varrho \rangle $, $S(\rho _{B})=S(\rho _{C})(=\ln 2)$.

From Lemmas 1 and 6, we have the following corollary.

{\bf Corollary 4.} A state of the form $(\lambda _{0},\pm \lambda _{1},\lambda
_{2},\lambda _{3},\lambda _{4})$ has the maximal von Neumann entanglement
entropy $S(\rho _{B})=S(\rho _{C})(=\ln 2)$ while $S(\rho _{A})<\ln 2$ if
and only if $\ $the state is $\lambda _{0}|000\rangle \pm \lambda
_{1}|100\rangle +\frac{1}{\sqrt{2}}|111\rangle $, where $\lambda _{1}\neq 0$.

{\bf Lemma 7.} The state $|\psi \rangle $\ of the form $(\lambda _{0},\pm \lambda
_{1},\lambda _{2},\lambda _{3},\lambda _{4})$ has the maximal von Neumann
entanglement entropy $S(\rho _{B})$($=\ln 2$) if and only if the state is
one of the following states 
\begin{eqnarray}
|\psi \rangle &=&\lambda _{0}|000\rangle \pm \lambda _{1}|100\rangle +\frac{1%
}{\sqrt{2}}|111\rangle , \\
|\psi \rangle &=&\frac{1}{\sqrt{2}}|000\rangle +\lambda _{3}|110\rangle
+\lambda _{4}|111\rangle , \\
\text{ }|\psi \rangle &=&\lambda _{0}|000\rangle -\frac{\lambda _{2}\lambda
_{4}}{\lambda _{3}}|100\rangle +\lambda _{2}|101\rangle +\lambda
_{3}|110\rangle +\lambda _{4}|111\rangle ,\text{ where } \\
\text{ }\lambda _{3} &>&\lambda _{2},2\lambda _{3}^{2}+2\lambda _{4}^{2}=1.
\end{eqnarray}

Proof. That $S(\rho _{B})=\ln 2$ implies 
\begin{equation}
\alpha _{B}=J_{1}+J_{3}+J_{4}=1/4  \label{al-b}
\end{equation}

From Eq. (\ref{al-b}), obtain

\begin{equation}
\lambda _{1}^{2}\lambda _{4}^{2}\pm 2\lambda _{1}\lambda _{2}\lambda
_{3}\lambda _{4}+\lambda _{2}^{2}\lambda _{3}^{2}+\lambda _{0}^{2}\lambda
_{3}^{2}+\lambda _{0}^{2}\lambda _{4}^{2}-1/4=0  \label{alph-b-1}
\end{equation}

Note that $\sum_{i=0}^{4}\lambda _{i}^{2}=1$. By substituting $\lambda
_{0}^{2}$ with $(1-(\lambda _{1}^{2}+\lambda _{2}^{2}+\lambda
_{3}^{2}+\lambda _{4}^{2}))$, from Eq. (\ref{alph-b-1}) obtain 
\begin{equation}
\allowbreak \lambda _{1}^{2}\lambda _{3}^{2}\pm 2\lambda _{1}\lambda
_{2}\lambda _{3}\lambda _{4}+\lambda _{2}^{2}\lambda _{4}^{2}+\lambda
_{3}^{4}+\allowbreak 2\lambda _{3}^{2}\lambda _{4}^{2}-\lambda
_{3}^{2}+\lambda _{4}^{4}-\lambda _{4}^{2}+\frac{1}{4}=0  \label{alph-b-2}
\end{equation}

Case 1 $\allowbreak \lambda _{3}=0$. Eq. (\ref{alph-b-2}) becomes 
\begin{equation}
\lambda _{2}^{2}\lambda _{4}^{2}+(\lambda _{4}^{2}-1/2)^{2}=0
\label{sol-b-1}
\end{equation}

From Eq. (\ref{sol-b-1}), obtain $\lambda _{2}=0$ and $\lambda _{4}^{2}=1/2$
and

\begin{equation}
|\psi \rangle =\lambda _{0}|000\rangle \pm \lambda _{1}|100\rangle +\frac{1}{%
\sqrt{2}}|111\rangle .  \label{state-b-1}
\end{equation}

It is known that $S(\rho _{B})=S(\rho _{C})=\ln 2$ for the state in Eq. (\ref%
{state-b-1}).

Case 2. $\lambda _{3}\neq 0$. \ $\allowbreak $

Case 2.1. $\lambda _{2}=0$. Eq. (\ref{alph-b-2}) becomes 
\begin{equation}
\lambda _{4}^{4}+\allowbreak 2\lambda _{3}^{2}\lambda _{4}^{2}-\lambda
_{4}^{2}+\allowbreak \lambda _{1}^{2}\lambda _{3}^{2}+\lambda
_{3}^{4}-\lambda _{3}^{2}+\frac{1}{4}=0.  \label{alph-b-3}
\end{equation}

$\allowbreak $The discriminant for $\lambda _{4}^{2}$ in Eq. (\ref{alph-b-3}%
) is 
\begin{equation}
\Delta =(\allowbreak 2\lambda _{3}^{2}-1)^{2}-4(\lambda _{1}^{2}\lambda
_{3}^{2}+\lambda _{3}^{4}-\lambda _{3}^{2}+\frac{1}{4})=\allowbreak
-4\lambda _{1}^{2}\lambda _{3}^{2}.\allowbreak  \label{delta-3}
\end{equation}

Let $\Delta =0$ in Eq. (\ref{delta-3}). Then, obtain $\lambda _{1}=0$. From
Eq. (\ref{alph-b-3}), obtain $\lambda _{4}^{2}=\allowbreak \frac{1}{2}%
-\lambda _{3}^{2}$.\ Then, $\lambda _{0}^{2}=1/2$. Thus, obtain

\begin{equation}
|\psi \rangle =\frac{1}{\sqrt{2}}|000\rangle +\lambda _{3}|110\rangle
+\lambda _{4}|111\rangle .  \label{state-b-2}
\end{equation}

It is known that $S(\rho _{A})=S(\rho _{B})=\ln 2$ for the state in Eq. (\ref%
{state-b-2}).

Case 2.2. $\lambda _{2}\neq 0$.\ To solve $\lambda _{1}$ from Eq. (\ref%
{alph-b-2}), let $\Delta $ be the discriminant for $\lambda _{1}$. Then, 
\begin{eqnarray}
\Delta &=&(\pm 2\lambda _{2}\lambda _{3}\lambda _{4})^{2}-4\lambda
_{3}^{2}(\lambda _{2}^{2}\lambda _{4}^{2}+\lambda _{3}^{4}+\allowbreak
2\lambda _{3}^{2}\lambda _{4}^{2}-\lambda _{3}^{2}+\lambda _{4}^{4}-\lambda
_{4}^{2}+\frac{1}{4}) \\
&=&-\lambda _{3}^{2}\left( 2\lambda _{3}^{2}+2\lambda _{4}^{2}-1\right) ^{2}.
\label{delta-2}
\end{eqnarray}

Let $\Delta =0$ in Eq. (\ref{delta-2}). Then, obtain $2\lambda
_{3}^{2}+2\lambda _{4}^{2}-1=0$ from Eq. (\ref{delta-2})\ and $\lambda
_{1}=\allowbreak \frac{\lambda _{2}\lambda _{4}}{\lambda _{3}}$ from Eq. (%
\ref{alph-b-2}) when the sign for the second term in Eq. (\ref{alph-b-2})
takes \textquotedblleft $-$\textquotedblright . Then, $\lambda _{0}^{2}=%
\frac{\lambda _{3}^{2}-\lambda _{2}^{2}}{2\lambda _{3}^{2}}$ from $%
\sum_{i=0}^{4}\lambda _{i}^{2}=1$. Under that $2\lambda _{3}^{2}+2\lambda
_{4}^{2}=1$, $\lambda _{3}>\lambda _{2}$, and $\lambda _{i}\neq 0$, $i=2,3$,
obtain

\begin{equation}
|\psi \rangle =\lambda _{0}|000\rangle -\frac{\lambda _{2}\lambda _{4}}{%
\lambda _{3}}|100\rangle +\lambda _{2}|101\rangle +\lambda _{3}|110\rangle
+\lambda _{4}|111\rangle .  \label{state-b-3}
\end{equation}

One can verify $S(\rho _{B})=\ln 2$ for $|\psi \rangle $ in Eq. (\ref%
{state-b-3}).

Therefore, if a state of the form $(\lambda _{0},\pm \lambda _{1},\lambda
_{2},\lambda _{3},\lambda _{4})$ has $s(\rho _{B})=\ln 2$ then the state
must be one of the states in Eqs. (\ref{state-b-1}, \ref{state-b-2}, \ref%
{state-b-3}). Conversely, for the states in Eqs. (\ref{state-b-1}, \ref%
{state-b-2}, \ref{state-b-3}), clearly $s(\rho _{B})=\ln 2.$

From Lemmas 4, 6, and 7, we have the following corollary.

{\bf Corollary 5.} A state of the form $(\lambda _{0},\pm \lambda _{1},\lambda
_{2},\lambda _{3},\lambda _{4})\ $has the maximal von Neumann entanglement
entropy $S(\rho _{B})(=\ln 2)$ while $S(\rho _{A})<\ln 2$ and $S(\rho
_{C})<\ln 2$ if and only if the state is $\lambda _{0}|000\rangle -\frac{%
\lambda _{2}\lambda _{4}}{\lambda _{3}}|100\rangle +\lambda _{2}|101\rangle
+\lambda _{3}|101\rangle +\lambda _{4}|111\rangle $, where $2\lambda
_{3}^{2}+2\lambda _{4}^{2}=1$, $\lambda _{3}>\lambda _{2}$, and $\lambda
_{i}\neq 0$, $i=2,3$.

{\bf Lemma 8.} The state of the form $(\lambda _{0},\pm \lambda _{1},\lambda
_{2},\lambda _{3},\lambda _{4})$ has the maximal von Neumann entanglement
entropy $S(\rho _{C})$ ($=\ln 2$) if and only if the state is one of the
following three states. 
\begin{eqnarray}
|\psi \rangle &=&\lambda _{0}|000\rangle +\pm \lambda _{1}|100\rangle +\frac{%
1}{\sqrt{2}}|111\rangle , \\
|\psi \rangle &=&\frac{1}{\sqrt{2}}|000\rangle +\lambda _{2}|101\rangle
+\lambda _{4}|111\rangle , \\
|\psi \rangle &=&\lambda _{0}|000\rangle -\allowbreak \frac{\lambda
_{3}\lambda _{4}}{\lambda _{2}}|100\rangle +\lambda _{2}|101\rangle +\lambda
_{3}|110\rangle +\lambda _{4}|111\rangle ,\text{where} \\
\lambda _{2} &>&\lambda _{3},2\lambda _{2}^{2}+2\lambda _{4}^{2}=1.
\end{eqnarray}

Proof. That $S(\rho _{C})=\ln 2$ implies 
\begin{equation}
\alpha _{C}=J_{1}+J_{2}+J_{4}=1/4  \label{al-c}
\end{equation}

From Eq. (\ref{al-c}), obtain

\begin{equation}
\lambda _{1}^{2}\lambda _{4}^{2}\pm 2\lambda _{1}\lambda _{2}\lambda
_{3}\lambda _{4}+\lambda _{2}^{2}\lambda _{3}^{2}+\lambda _{0}^{2}\lambda
_{2}^{2}+\lambda _{0}^{2}\lambda _{4}^{2}-1/4=0  \label{alph-c-1}
\end{equation}

Note that $\sum_{i=0}^{4}\lambda _{i}^{2}=1$.\ Substituting $\lambda
_{0}^{2} $ with $(1-(\lambda _{1}^{2}+\lambda _{2}^{2}+\lambda
_{3}^{2}+\lambda _{4}^{2}))$ in Eq. (\ref{alph-c-1}), obtain 
\begin{equation}
\allowbreak \lambda _{1}^{2}\lambda _{2}^{2}\pm 2\lambda _{1}\lambda
_{2}\lambda _{3}\lambda _{4}+\lambda _{2}^{4}+2\lambda _{2}^{2}\lambda
_{4}^{2}-\allowbreak \lambda _{2}^{2}+\lambda _{3}^{2}\lambda
_{4}^{2}+\lambda _{4}^{4}-\lambda _{4}^{2}+\frac{1}{4}=0  \label{alph-c-2}
\end{equation}

Case 1. $\allowbreak \lambda _{2}=0$.

Eq. (\ref{alph-c-2}) becomes 
\begin{equation}
\lambda _{3}^{2}\lambda _{4}^{2}+(\lambda _{4}^{2}-1/2)^{2}=0
\label{alph-c-3}
\end{equation}

$\allowbreak $Then, $\lambda _{3}=0$ and $\lambda _{4}^{2}=1/2$ from Eq. (%
\ref{alph-c-3}). Thus, 
\begin{equation}
|\psi \rangle =\lambda _{0}|000\rangle \pm \lambda _{1}|100\rangle +\frac{1}{%
\sqrt{2}}|111\rangle .  \label{state-c-1}
\end{equation}

It is known that $S(\rho _{B})=S(\rho _{C})=\ln 2$ for the state in Eq. (\ref%
{state-c-1}).

$\allowbreak $Case 2. $\lambda _{2}\neq 0$.

Case 2.1. $\lambda _{3}=0$. Eq. (\ref{alph-c-2}) becomes 
\begin{equation}
\lambda _{4}^{4}+2\lambda _{2}^{2}\lambda _{4}^{2}-\lambda
_{4}^{2}+\allowbreak \lambda _{1}^{2}\lambda _{2}^{2}+\lambda
_{2}^{4}-\allowbreak \lambda _{2}^{2}+\frac{1}{4}=0  \label{alph-c-4}
\end{equation}

To solve $\lambda _{4}^{2}$ from Eq. (\ref{alph-c-4}), let $\Delta $ be the
discriminant in Eq. (\ref{alph-c-4}). Then,

\begin{equation}
\Delta =(2\lambda _{2}^{2}-1)^{2}-4(\allowbreak \lambda _{1}^{2}\lambda
_{2}^{2}+\lambda _{2}^{4}-\allowbreak \lambda _{2}^{2}+\frac{1}{4}%
)=-4\lambda _{1}^{2}\lambda _{2}^{2}.  \label{delta-c-3}
\end{equation}

Let $\Delta =0$ in Eq. (\ref{delta-c-3}). Then, obtain $\lambda _{1}=0$ from
Eq. (\ref{delta-c-3}). From Eq. (\ref{alph-c-4}), $\lambda
_{4}^{2}=\allowbreak \frac{1}{2}-\lambda _{2}^{2}$. Then, $\lambda _{0}=1/%
\sqrt{2}$. Thus, obtain 
\begin{equation}
|\psi \rangle =\frac{1}{\sqrt{2}}|000\rangle +\lambda _{2}|101\rangle
+\lambda _{4}|111\rangle .  \label{state-c-2}
\end{equation}

It is known that $S(\rho _{A})=S(\rho _{C})=\ln 2$ for the state in Eq. (\ref%
{state-c-2}).

Case 2.2. $\lambda _{3}\neq 0$. To solve $\lambda _{1}$ from Eq. (\ref%
{alph-c-2}), \ let $\Delta $\ be the discriminant for $\lambda _{1}$ in Eq. (%
\ref{alph-c-2}). Then,

\begin{eqnarray}
\Delta &=&(\pm 2\lambda _{2}\lambda _{3}\lambda _{4})^{2}-4\lambda
_{2}^{2}(\lambda _{2}^{4}+2\lambda _{2}^{2}\lambda _{4}^{2}-\allowbreak
\lambda _{2}^{2}+\lambda _{3}^{2}\lambda _{4}^{2}+\lambda _{4}^{4}-\lambda
_{4}^{2}+\frac{1}{4})  \label{delta-c-1} \\
&=&-\lambda _{2}^{2}\left( 2\lambda _{2}^{2}+2\lambda _{4}^{2}-1\right)
^{2}.\allowbreak  \label{delta-c-4}
\end{eqnarray}

$\allowbreak $Let $\Delta =0$ in Eq. (\ref{delta-c-4}). Then, obtain $%
2\lambda _{2}^{2}+2\lambda _{4}^{2}-1=0$. From Eq. (\ref{alph-c-2}), obtain $%
\lambda _{1}=\allowbreak \frac{\lambda _{3}\lambda _{4}}{\lambda _{2}}$ when
the sign for the second term in Eq. (\ref{alph-c-2}) takes \textquotedblleft 
$-$\textquotedblright . Then, $\lambda _{0}^{2}=\allowbreak \frac{1}{%
2\lambda _{2}^{2}}\left( \lambda _{2}^{2}-\lambda _{3}^{2}\right) $. So,
under that $\lambda _{2}>\lambda _{3}$, $2\lambda _{2}^{2}+2\lambda
_{4}^{2}=1$, $\lambda _{i}\neq 0$, $i=2,3$, obtain 
\begin{equation}
|\psi \rangle =\lambda _{0}|000\rangle -\allowbreak \frac{\lambda
_{3}\lambda _{4}}{\lambda _{2}}|100\rangle +\lambda _{2}|101\rangle +\lambda
_{3}|110\rangle +\lambda _{4}|111\rangle .  \label{state-c-3}
\end{equation}

One can verify that $S(\rho _{C})=\ln 2$ for $|\psi \rangle $ in Eq. (\ref%
{state-c-3}).

Therefore, if a state of the form $(\lambda _{0},\pm \lambda _{1},\lambda
_{2},\lambda _{3},\lambda _{4})$ has $s(\rho _{C})=\ln 2$ then the state
must be one of the states in Eqs. (\ref{state-c-1}, \ref{state-c-2}, \ref%
{state-c-3}). Conversely, for the states in Eqs. (\ref{state-c-1}, \ref%
{state-c-2}, \ref{state-c-3}), clearly $s(\rho _{C})=\ln 2$.

Lemmas 3, 6, 8 imply the following corollary.

{\bf Corollary 6.} A state $|\psi \rangle $ of the form $(\lambda _{0},\pm \lambda
_{1},\lambda _{2},\lambda _{3},\lambda _{4})$\ has the maximal von Neumann
entanglement entropy $S(\rho _{C})(=\ln 2)$ while $S(\rho _{A})<\ln 2$ and $%
S(\rho _{B})<\ln 2$ if and only if the state is $\lambda _{0}|000\rangle
-\allowbreak \frac{\lambda _{3}\lambda _{4}}{\lambda _{2}}|100\rangle
+\lambda _{2}|101\rangle +\lambda _{3}|101\rangle +\lambda _{4}|111\rangle .$%
, where $\lambda _{2}>\lambda _{3}$, $2\lambda _{2}^{2}+2\lambda _{4}^{2}=1$%
, $\lambda _{i}\neq 0$, $i=2,3$.

\section{Propositions used in Section \ref{sec:maxvon}}
\label{app:D}

{\bf Proposition 4.} $\eta _{0}^{2}+\eta _{1}^{2}+\eta _{2}^{2}+\eta _{3}^{2}+\eta
_{4}^{2}=(p_{1})^{2}+(p_{2})^{2}+(p_{3})^{2}+q_{0}^{2}$.

Proof. We consider the case $p^{1}p^{2}p^{3}q_{0}<0$. A complicated
calculation yields the following.

\begin{eqnarray}
\eta _{0}^{2} &=&\frac{(p^{2}p^{3}-p^{1}q_{0})(p^{1}p^{3}-p^{2}q_{0})}{%
p^{1}p^{2}-p^{3}q_{0}},  \label{ata-0} \\
\eta _{1}^{2} &=&-\frac{%
p^{1}p^{2}p^{3}q_{0}((p^{1})^{2}+(p^{2})^{2}-(p^{3})^{2}-q_{0}^{2})^{2}}{%
(p^{2}p^{3}-p^{1}q_{0})(p^{1}p^{3}-p^{2}q_{0})(p^{1}p^{2}-p^{3}q_{0})},
\label{ata-1} \\
\eta _{2}^{2} &=&\frac{(p^{1}p^{2}-p^{3}q_{0})\left(
p^{1}p^{3}+p^{2}q_{0}\right) ^{2}}{%
(p^{2}p^{3}-p^{1}q_{0})(p^{1}p^{3}-p^{2}q_{0})},  \label{ata-2} \\
\eta _{3}^{2} &=&\frac{(p^{1}p^{2}-p^{3}q_{0})\left(
p^{2}p^{3}+p^{1}q_{0}\right) ^{2}}{%
(p^{2}p^{3}-p^{1}q_{0})(p^{1}p^{3}-p^{2}q_{0})},  \label{ata-3} \\
\eta _{4}^{2} &=&-\frac{4p^{1}p^{2}p^{3}q_{0}\left(
p^{1}p^{2}-p^{3}q_{0}\right) }{(p^{2}p^{3}-p^{1}q_{0})(p^{1}p^{3}-p^{2}q_{0})%
}.  \label{ata-4}
\end{eqnarray}

Via Eqs. (\ref{ata-0}-\ref{ata-4}), it is easy to verify that $\eta
_{0}^{2}+\eta _{1}^{2}+\eta _{2}^{2}+\eta _{3}^{2}+\eta
_{4}^{2}=(p_{1})^{2}+(p_{2})^{2}+(p_{3})^{2}+q_{0}^{2}$.

{\bf Proposition 5.} $\frac{\eta _{0}}{\sqrt{%
(p^{1})^{2}+(p^{2})^{2}+(p^{3})^{2}+q_{0}^{2}}}=\frac{1}{\sqrt{2}}$ if and
only if $\allowbreak p^{1}p^{2}+q_{0}p^{3}=0\allowbreak $ or $%
(p^{1})^{2}+(p^{2})^{2}-(p^{3})^{2}-q_{0}^{2}=0$.

Proof. Via Eq. (\ref{ata-0}), $\eta _{0}^{2}=\frac{%
(p^{2}p^{3}-p^{1}q_{0})(p^{1}p^{3}-p^{2}q_{0})}{p^{1}p^{2}-p^{3}q_{0}}$.
Then, 
\begin{equation}
\frac{\eta _{0}}{\sqrt{(p^{1})^{2}+(p^{2})^{2}+(p^{3})^{2}+q_{0}^{2}}}=\sqrt{%
\frac{(p^{2}p^{3}-p^{1}q_{0})(p^{1}p^{3}-p^{2}q_{0})}{%
(p^{1}p^{2}-p^{3}q_{0})((p^{1})^{2}+(p^{2})^{2}+(p^{3})^{2}+q_{0}^{2})}}.
\label{prop-2}
\end{equation}

Assume $\frac{\eta _{0}}{\sqrt{(p^{1})^{2}+(p^{2})^{2}+(p^{3})^{2}+q_{0}^{2}}%
}=\frac{1}{\sqrt{2}}$. Then, via factoring, a calculation yields 
\begin{eqnarray}
&&((p^{1})^{2}+(p^{2})^{2}+(p^{3})^{2}+q_{0}^{2})(p^{1}p^{2}-p^{3}q_{0}) 
\nonumber \\
&&-2(p^{2}p^{3}-p^{1}q_{0})(p^{1}p^{3}-p^{2}q_{0})  \nonumber \\
&=&\allowbreak \left( p^{1}p^{2}+q_{0}p^{3}\right)
((p^{1})^{2}+(p^{2})^{2}-(p^{3})^{2}-q_{0}^{2})
\end{eqnarray}

Therefore, $\allowbreak p^{1}p^{2}+q_{0}p^{3}=0\allowbreak $ or $%
(p^{1})^{2}+(p^{2})^{2}-(p^{3})^{2}-q_{0}^{2}=0$.

Conversely, we next show if $\allowbreak p^{1}p^{2}+q_{0}p^{3}=0\allowbreak $
or $(p^{1})^{2}+(p^{2})^{2}-(p^{3})^{2}-q_{0}^{2}=0$, then $\frac{\eta _{0}}{%
\sqrt{(p^{1})^{2}+(p^{2})^{2}+(p^{3})^{2}+q_{0}^{2}}}=\frac{1}{\sqrt{2}}$.

Case 1. Assume that $p^{1}p^{2}+q_{0}p^{3}=0\allowbreak $. A calculation
yields

\begin{eqnarray}
&&(p^{2}p^{3}-p^{1}q_{0})(p^{1}p^{3}-p^{2}q_{0}) \\
&=&p^{1}p^{2}q_{0}^{2}+p^{1}p^{2}(p^{3})^{2}-((p^{1})^{2}+(p^{2})^{2})q_{0}p^{3}
\\
&=&p^{1}p^{2}q_{0}^{2}+p^{1}p^{2}(p^{3})^{2}-((p^{1})^{2}+(p^{2})^{2})(-p^{1}p^{2})
\\
&=&p^{1}p^{2}((p_{1})^{2}+(p_{2})^{2}+(p_{3})^{2}+q_{0}^{2})
\end{eqnarray}

and 
\begin{eqnarray}
&&(p^{1}p^{2}-p^{3}q_{0})((p^{1})^{2}+(p^{2})^{2}+(p^{3})^{2}+q_{0}^{2}) \\
&=&2p^{1}p^{2}((p^{1})^{2}+(p^{2})^{2}+(p^{3})^{2}+q_{0}^{2})
\end{eqnarray}

Therefore, from Eq. (\ref{prop-2}), $\frac{\eta _{0}}{\sqrt{%
(p^{1})^{2}+(p^{2})^{2}+(p^{3})^{2}+q_{0}^{2}}}=\frac{1}{\sqrt{2}}$.

Case 2. Assume 
\begin{equation}
(p^{1})^{2}+(p^{2})^{2}-(p^{3})^{2}-q_{0}^{2}=0.  \label{hyp-2}
\end{equation}

From Eq. (\ref{hyp-2}), obtain $%
(p^{3})^{2}+q_{0}^{2}=(p^{1})^{2}+(p^{2})^{2} $ and then 
\begin{eqnarray}
&&(p^{1}p^{2}-p^{3}q_{0})((p^{1})^{2}+(p^{2})^{2}+(p^{3})^{2}+q_{0}^{2}) \\
&=&2(p^{1}p^{2}-p^{3}q_{0})((p^{1})^{2}+(p^{2})^{2}),
\end{eqnarray}%
and

\begin{eqnarray}
&&(p^{2}p^{3}-p^{1}q_{0})(p^{1}p^{3}-p^{2}q_{0}) \\
&=&-(p^{1})^{2}q_{0}p^{3}+p^{1}p^{2}(q_{0}^{2}+(p^{3})^{2})-(p^{2})^{2}q_{0}p^{3}
\\
&=&-(p^{1})^{2}q_{0}p^{3}+p^{1}p^{2}((p^{1})^{2}+(p^{2})^{2})-(p^{2})^{2}q_{0}p^{3}
\\
&=&(p^{1}p^{2}-p^{3}q_{0})((p^{1})^{2}+(p^{2})^{2}).
\end{eqnarray}

Therefore, from Eq. (\ref{prop-2}), $\frac{\eta _{0}}{\sqrt{%
(p^{1})^{2}+(p^{2})^{2}+(p^{3})^{2}+q_{0}^{2}}}=\frac{1}{\sqrt{2}}$.

\bigskip

\end{appendices}

\end{document}